\documentclass[a4paper,11pt]{article}
\pdfoutput=1 

\usepackage{jcappub} 

\usepackage[T1]{fontenc} 

\usepackage{graphicx}
\usepackage{float}
\usepackage{hyperref}
\usepackage{xcolor, xstring}
\usepackage{amssymb, amsmath, mathtools}
\usepackage{listings}
\usepackage{lscape}
\usepackage{lipsum}
\usepackage{multicol}
\usepackage{physics}
\usepackage{natbib}
\usepackage{txfonts}
\usepackage{xspace}
\usepackage{comment}
\usepackage[normalem]{ulem} 

\newcommand{\pot}[2]{#1 \times 10^{#2}}

\newcommand{\boostO}{\hat{\mathcal{O}}_x}

\newcommand{\KompO}{\hat{\mathcal{K}}_x}

\newcommand{\DiffO}{\hat{\mathcal{D}}_x}
\newcommand{\DiffstarO}{\hat{\mathcal{D}}^*_x}
\newcommand{\DiffOp}{\hat{\mathcal{D}}_{x'}}

\newcommand{\diff}{\mathop{}\!\mathrm{d}}

\newcommand{\nbb}{n_{\rm bb}}
\newcommand{\Gspec}{{G}}
\newcommand{\Yspec}{{{Y}}}
\newcommand{\Ynspec}[1]{{Y}_{#1}}
\newcommand{\Mspec}{{M}}

\newcommand{\COBEF}{{\it COBE/FIRAS}\xspace}

\newcommand{\expf}[1]{{{\rm e}^{#1}}}

\newcommand{\cm}{{\rm cm}}

\newcommand{\Tz}{{T_{z}}}

\newcommand{\vbeta}{{{\boldsymbol\beta}}}
\newcommand{\vbetah}{{\hat{{\boldsymbol\beta}}}}

\newcommand{\vgh}{{\hat{\boldsymbol\gamma}}}

\newcommand{\kD}{k_{\rm D}}

\newcommand{\xc}{x_{\rm c}}

\newcommand{\id}{{\,\rm d}}

\newcommand{\beq}{\begin{equation}}   %

\newcommand{\eeq}{\end{equation}}   %

\newcommand{\beqa}{\begin{eqnarray}}   %

\newcommand{\eeqa}{\end{eqnarray}}   %

\newcommand{\bealf}[1]{\begin{align} #1 \end{align}}

\newcommand{\beal}{\begin{align}}
\newcommand{\enal}{\end{align}}

\newcommand{\bspl}{\begin{split}}

\newcommand{\espl}{\end{split}}

\newcommand{\bsub}{\begin{subequations}}

\newcommand{\esub}{\end{subequations}}

\newcommand{\bmulti}{\begin{multline}}   %

\newcommand{\beqm}{\begin{mathletters}}   %

\newcommand{\eeqm}{\end{mathletters}}   %

\newcommand{\kB}{k_{\rm B}}

\newcommand{\me}{m_{\rm e}}

\newcommand{\Ne}{N_{\rm e}}

\newcommand{\Te}{T_{\rm e}}

\newcommand{\Tg}{T_{\gamma}}

\newcommand{\The}{\theta_{\rm e}}

\newcommand{\Thg}{\theta_{\gamma}}

\newcommand{\sigT}{\sigma_{\rm T}}

\newcommand{\vek} [1]{\mbox{\boldmath${#1}$\unboldmath}}

\newcommand{\Thz}{\theta_{z}}

\newcommand{\ysc}{y_{z}}

\usepackage{bbm}
\def\i{\mathbbm{i}}

\title{Spectro-spatial evolution of the CMB II: generalised Boltzmann hierarchy}

\author[a]{Jens Chluba}

\author[b,c]{, Andrea Ravenni}

\author[a]{and Thomas Kite}

\affiliation[a]{Jodrell Bank Centre for Astrophysics, School of Physics and Astronomy, The University of Manchester, Oxford Road, Manchester, M13 9PL, U.K.}

\affiliation[b]{Dipartimento di Fisica e Astronomia Galileo Galilei, Universita degli Studi di Padova, via Marzolo 8, I-35131, Padova, Italy.}
\affiliation[c]{INFN, Sezione di Padova, via Marzolo 8, I-35131, Padova, Italy.}

\emailAdd{Jens.Chluba@Manchester.ac.uk}
\emailAdd{Andrea.Ravenni@unipd.it}
\emailAdd{Thomas.Kite@Manchester.ac.uk}

\date{Aug 2022}

\begin{document}

\abstract{In this paper, we formulate a generalised photon Boltzmann hierarchy that allows us to model the evolution and creation of spectral distortion anisotropies in the early Universe. We directly build on our first paper in this series, extending the thermalisation Green's function treatment to the anisotropic case.
We show that the problem can be described with the common Boltzmann hierarchy for the photon field extended by new spectral parameters -- a step that reduces the complexity of the calculation by at least two orders of magnitude. Our formalism describes the effects of i) Doppler and potential driving, ii) spectral evolution by Compton scattering, iii) perturbed thermalisation and iv) anisotropic heating on the distortion anisotropies.
We highlight some of the main physical properties of the equations and also outline the steps for computing CMB power spectra including distortion anisotropies. Limitations and extensions of the formulation are also briefly discussed.
The novel Boltzmann hierarchy given here is the basis for a series of companion papers studying how distortion anisotropies evolve in the perturbed Universe and which physical processes could be constrained using future CMB imaging techniques.}

\maketitle
\flushbottom

\newpage
\section{Introduction}
\vspace{-2mm}
The cosmic microwave background (CMB) has been a goldmine for furthering our understanding of the cosmos \citep{COBE4yr, WMAP_params, ACT11, spt, Planck2015params}. We are now entering a new phase, in which novel cosmological observables are moving to the focus of our efforts \citep{Abazajian2016S4SB, Delabrouille2018, PICO, Delabrouille2021}. One of these observables is spectral distortions (SDs) of the CMB \citep{Chluba2019BAAS, Chluba2021ExA}.
In this work, we develop a novel Boltzmann formulation that can describe the evolution of SD {\it anisotropies} in the early Universe.  
We base the treatment on our first paper in the series \citep[][henceforth referred to as paper I]{Chluba2022TFHI}, which introduced a novel discretisation of the thermalisation Green's function to efficiently describe the evolution of average SDs. In paper I, we were limited to running one single thermalisation history at the time, effectively obtaining a solution for the average distortion intensity spectrum, $\Delta I_\nu(t)$, at a given time $t$ and frequency $\nu$. In simple words, we now deliver the tools to repeat the calculation along multiple lines of sight and including the effect of perturbations on the CMB signal evolution. 
This will open the way to studying SD physics using standard methods known from CMB imaging, without the need for absolute calibration.

We refer the reader to paper I for more introduction and motivation to the topic. The steps taken here explain in detail how to extend the thermalisation Green's function treatment to the anisotropic case. We are in particular keen on including thermalisation effects to the evolution. This captures the full spectral evolution due to Compton scattering, double Compton (DC) and Bremsstrahlung (BR), which are so crucial in the formation of SDs at redshifts $z\gtrsim 10^4$ \citep{Sunyaev1970mu, Danese1982, Burigana1991, Hu1993, Chluba2014}. These effects can change the type of the distortion and gradually {\it rotate} $y$-type distortion sources into $\mu$ and ultimately thermalise the distortion completely. 
The rotation is with respect to the energy in various spectral components, which by construction is conserved across the spectral basis.
A perturbative formulation that includes some of the Doppler boosting effects has been given in \cite{Khatri2012mix} and \cite{Ota2017, Haga2018}; however, this does not capture the effects of repeated Compton scattering at $z>10^5$ and also does not describe the conversion of $\mu$ to $T$, thus having limited applicability, which we overcome here.

In Sect.~\ref{sec:Boltzmann-formulation}, we start by recapping the main outcomes of paper I. We then move on to describing some of the distortion effect in the Thomson limit of the kinetic equation. This only captures the effect of Doppler and potential driving as well as free-streaming mixing on the SD signals across the sky, but provides valuable insight preparing for the general case.
In Sect.~\ref{sec:Comp_effects}, we add thermalisation terms to complete our treatment. Details of the derivation are giving in Appendix~\ref{app:collision_first} to \ref{app:approximations_kin}, which we recommend for in depth reading.
The tools for computing the CMB power spectra for various combinations of temperature and SD parameters are given in Sect.~\ref{eq:tools}, and a discussion of the basic expectations is presented in Sect.~\ref{sec:understanding}. Numerical solutions and first forecasts will be presented in paper III (in preparation). In Sect.~\ref{sec:discussion}, we present our conclusions and also highlight some of the limitation of the formulation and further work that may become important.

\vspace{-2mm}
\section{Extended Boltzmann hierarchy with primordial spectral distortions}
\label{sec:Boltzmann-formulation}
\vspace{-2mm}
The goal of this section is the provide an approximate description of the spectro-spatial thermalisation problem at first order in perturbation theory. 
This greatly generalizes previous treatments of the problem allowing us to capture the main sources of distortions in the presence of perturbations. 
Unlike the usual split, here we have two small parameters to work with, one for the average energy release, $\epsilon_\gamma=\Delta \rho_\gamma/\rho_\gamma$, and one for the primordial curvature perturbations, $\zeta$. We keep terms up to $\mathcal{O}(\epsilon_\gamma \zeta)$, thereby allowing to capture all linear order (spatial) perturbation effects.

We will use the approach presented in paper I to model the evolution of the local monopole across the chosen spectral basis, while the spatial evolution is treated using the standard multipole decomposition which includes the effect of Thomson scattering and free-streaming.
We will make direct use of the main results from paper I and refer the interested reader to that paper for clarification of the notation and details of the Green's function discretisation and choice of basis functions.

\subsection{Paper I and zeroth order problem}
In paper I, we essentially solved the zeroth order photon equation for the evolution of the average CMB spectrum. This problem can be described with the kinetic equation \citep[see Sect.~4.1.1 of][]{Chluba:2x2}:
\begin{align}
\label{eq:zeroth_equation}
\frac{1}{\dot{\tau}}\,\frac{\partial \Delta n^{(0)}}{\partial t}
\approx \Thz \Theta^{(0)}_{\rm eq}\,\Yspec+\Thz\,\KompO\,\Delta n^{(0)}
- \frac{\Lambda\,(1-\expf{-x})}{x^3}\Delta n^{(0)}
+
\frac{\Lambda}{x^2}\,\nbb\, \Theta^{(0)}_{\rm eq}
+\frac{\dot{\mathcal{Q}}^{(0)}}{\dot{\tau}}\,\Yspec.
\end{align}
for the evolution of the photon occupation number distortion $\Delta n^{(0)}\equiv \Delta n^{(0)}(t, x)$, defined with respect to the blackbody, $\nbb(x=h\nu/\kB\Tz)=1/(\expf{x}-1)$ at a temperature $\Tz$.
Here, the dot denotes time derivatives and we introduced the Thomson optical depth, $\tau = \int \sigT \Ne c \id t$, which is evaluated at the background level assuming the standard recombination history from {\tt CosmoRec} \cite{Chluba2010b}. 
The temperature variables are presented as $\theta_i=\kB T_i/\me c^2$, with $\Thz \propto \Tz \propto (1+z)$.  The Kompaneets operator is denoted by $\KompO=x^{-2} \partial_x x^4 \partial_x + x^{-2} \partial_x x^4 A(x)$, with $A(x)=1+2\nbb(x)$, and $\Lambda=\Lambda(x, \Thz)$ determines the photon production rate by double Compton (DC) and Bremsstrahlung (BR). These can be computed accurately using {\tt BRpack} \citep{BRpack} and {\tt DCpack} \citep{Ravenni2020DC}.
The electron temperature perturbation, $\Theta^{(0)}_{\rm eq}=\Delta \Te^{(0)}/\Tz$, and effective heating rate, $\dot{\mathcal{Q}}^{(0)}$, are given by
\bealf{
\label{app:Theta_eq}
\Theta^{(i)}_{\rm eq}
&=\frac{\int x^3 \Delta n^{(i)}_0 \, w_y \id x}{4E_{\nbb}}
\qquad {\rm and}\qquad
\dot{\mathcal{Q}}^{(0)}\equiv \frac{\dot{Q}^{(0)}_{\rm c}}{\rho_z},
}
with $y$-weight factor $w_y= \Yspec/\Gspec=x A(x)-4=x\frac{\expf{x}+1}{\expf{x}-1}-4$ and where $E_f=\int x^3 f(x) \id x$ is the energy density integral of $f(x)$. The expression for $\Theta^{(i)}_{\rm eq}$ can be obtained by balancing Compton heating and cooling (Appendix~\ref{app:Compton_energy_exchange}). The distortion sources from heating relate directly to $\dot{Q}^{(0)}_{\rm c}$ from collisions (see Appendix~\ref{app:heating_Te}). In equation~\eqref{app:Theta_eq}, $\rho_z=\frac{8\pi h}{c^3}\left(\frac{\kB\Tz}{h}\right)^4\,E_{\nbb}\approx 0.261\,{\rm eV/\cm^3}\,[T_0(1+z)/2.726\,{\rm K}]^4$ is the energy density of a blackbody at temperature $\Tz$.

Usually, Eq.~\eqref{eq:zeroth_equation} is solved on a frequency-grid for the spectral distortion, $\Delta n^{(0)}(t, x)$, e.g., using {\tt CosmoTherm} \citep{Chluba2011therm}. This task can become very time-consuming and certainly is not easily extendable to anisotropic distortions. The main result of paper I was to demonstrate that the problem can be approximately written as a matrix equation. The derivation uses the Ansatz
$\Delta n^{(0)}(t, x)\approx \vek{B}(x) \cdot \vek{y}^{(0)}(t)$ to discretise the average photon spectrum. Here, $\vek{B}=(\Gspec(x),\Yspec(x), Y_1(x),\ldots,Y_N(n),\Mspec(x))^T$ denotes the computation basis, based on the standard distortion shapes, $G, Y$ and $M$ as well as the boosted signals, $Y_k=(1/4)^k \boostO^k Y$ with boost generator, $\boostO=-x\partial_x$. This then yields the evolution equation
\begin{align}
\label{eq:Main_Eq_photon_injection}
&\quad\quad\quad\quad\quad\quad
\frac{\partial \vek{y}^{(0)}}{\partial t}
\approx \dot{\tau}\Thz \left[M_{\rm K}\,\vek{y}^{(0)}
+\vek{D}^{(0)}\right]+\frac{\dot{\vek{Q}}^{(0)}}{4},
\nonumber
\\
\vek{D}^{(0)}
&=
\left(
     \gamma_T\xc\,\mu^{(0)}, 0, 0, \ldots, 0, -\gamma_N\xc\,\mu^{(0)}
\right)^T,\qquad
\dot{\vek{Q}}^{(0)}
=\left(0, \dot{\mathcal{Q}}^{(0)}, 0, \ldots, 0, 0\right)^T.
\end{align}
for the spectral parameters, $\vek{y}^{(0)}(t)$. Here, $\xc$ is the critical frequency for photon emission, and the emission coefficients are $\gamma_T\approx 0.1387$ and $\gamma_N\approx 0.7769$. The Kompaneets mixing matrix, $M_K$ describes the (photon number- and energy-conserving) rotation of the spectral parameters in each scattering, and is pre-computed for different basis sizes. From Eq.~\eqref{eq:Main_Eq_photon_injection} one can already anticipate the main thermalisation terms; however, a few augmentations will become important as we show now. 
We also note, that for Eq.~\eqref{eq:Main_Eq_photon_injection}, the problem was linearised respect to the distortion.
Possible sources of distortions from photon injection processes \citep[e.g.,][]{Chluba2015GreensII, Bolliet2020PI} were also neglected.

\subsection{General statement of the problem in first order of perturbation theory}
At first order in perturbation theory, the evolution equation for the anisotropic photon occupation number, $n^{(1)}=n^{(1)}(t, x, \vek{r}, \vgh)$, at location $\vek{r}$ and in the direction $\vgh$ reads  \citep[e.g.,][]{Ma1995, Hu1996, DodelsonBook}
\bealf{
\label{eq:evol_1}
&\frac{\partial n^{(1)}}{\partial t}+\frac{c\vgh}{a}\cdot \nabla n^{(1)}+
\boostO n^{(0)} \left(\frac{\partial \Phi^{(1)}}{\partial t}+ \frac{c\vgh}{a}\cdot \nabla\Psi^{(1)} \right)=\mathcal{C}^{(1)}[n].
}
Here, we work with the metric perturbations in conformal Newtonian gauge in which the line element is $\diff s^2 = a^2(-\expf{2\Psi} \diff \eta^2 + \expf{2\Phi} \delta_{ij} \diff x^i \diff x^j)$, neglecting vector and tensor perturbations; $\mathcal{C}^{(1)}[n]$ denotes the rather complicated collision term \citep[e.g.,][]{Hu1994pert, Senatore2009, Chluba:2x2}, that accounts for the effect of Thomson scattering and thermalisation processes. Usually, one would neglect all distortions or thermalisation effects, such that $\boostO n^{(0)}\approx \boostO \nbb \approx \Gspec(x)$. With the Ansatz $n^{(1)}\approx \Theta^{(1)}\,\Gspec(x)$, where $\Theta^{(1)}=\Delta T/T_0$ describes the fractional CMB temperature perturbation, this leads to the standard brightness temperature equations in the Thomson limit \citep{DodelsonBook, Ma1995, Hu1997}:
\bealf{
\label{eq:evol_1_Theta_T}
&\frac{\partial \Theta^{(1)}}{\partial t}+\frac{c\vgh}{a}\cdot \nabla \Theta^{(1)}+
\frac{\partial \Phi^{(1)}}{\partial t}+ \frac{c\vgh}{a}\cdot \nabla\Psi^{(1)} =\dot{\tau}\left[\Theta^{(1)}_0+\frac{1}{10}\,\Theta^{(1)}_2 - \Theta^{(1)} + \beta^{(1)}\chi\right],
}
where $\Theta^{(1)}_\ell$ is the Legendre transform of the photon temperature field, $\Phi^{(1)}$ and $\Psi^{(1)}$ are the potential perturbations, $\beta^{(1)}$ the baryon speed, and $\chi=\vbetah\cdot \vgh$ is the direction cosine between the baryon velocity $\vbeta$ and the photon direction, $\vgh$. Polarisation terms were not included here, but do not affect the main arguments presented below. For our numerical solutions, they are added back as usual \citep{Ma1995}.
Equation~\eqref{eq:evol_1_Theta_T} will now be generalized to include the effect of a non-vanishing average distortion, spectral evolution, anisotropic heating and perturbed thermalisation. These effects lead to small corrections to the brightness temperature equations but give the leading order terms in the distortion hierarchy.

Accounting for all the thermalisation effects self-consistently is beyond the scope of this paper. However, if we assume that locally thermalisation is {\it only} mediated through the monopole part of the spectrum, the situation becomes more tractable. This is in fact well-motivated since for the anisotropies the much faster Thomson scattering process dominates, while Thomson terms are absent in the monopole, making thermalisation terms dominant there. We will furthermore not attempt to solve for higher order corrections to the CMB temperature anisotropy field, as this would entail a full treatment in second order perturbation theory \citep[e.g.,][]{Bartolo2006, Pitrou2009, Senatore2009} including additional modifications to more correctly treat distortion terms. 

To better appreciate the various new effects we will proceed in a step-by-step manner first only considering Thomson scattering terms, $\propto \dot{\tau}$. These will reveal the optimal basis given an average distortion and also already show the main effects in terms of distortion anisotropy generation. The spectral evolution across the various distortion types is then included using the modified Green's function treatment with the spectral mixing occurring only in the local monopole. The scattering efficiency is $\propto \dot{\tau} \Thz$ and hence suppressed relative to the Thomson terms. However, at $z\gtrsim 10^4$, we expected spectral evolution to change the distortion signals in an observable way.

We give many of the derivation details in Appendix~\ref{app:collision_first}. In addition, we will consider various limiting solution in Appendix~\ref{app:limits_evol}. In Appendix~\ref{app:approximations_kin}, we furthermore explicitly carry out some of the approximations that are required to simplify the problem to the expressions given in the main text.
We refer the interested reader to these sections for an in depth understanding. 
However, we hope that the result presented in this section can be mostly understood without the detailed (and sometimes cumbersome) derivations.

\subsection{Effect of Doppler and potential terms in the Thomson limit}
As the first and simplest scenario, let us consider the case where heating happened well before the recombination era and the average distortion is long frozen into its final state. We shall also assume that any thermalisation effects and energy exchange terms can be neglected. Thomson scattering isotropises the medium and leads to damping of the perturbations at small scales. This picture is valid for modes that become dynamic at $z\lesssim 10^4$. The photon evolution equation then reads 
\bealf{
\label{eq:evol_1_lim_1}
&\frac{\partial n^{(1)}}{\partial t}+\frac{c\vgh}{a}\cdot \nabla n^{(1)}+
\boostO n^{(0)} \left(\frac{\partial \Phi^{(1)}}{\partial t}+ \frac{c\vgh}{a}\cdot \nabla\Psi^{(1)} \right)\approx \dot\tau\left[n^{(1)}_0+\frac{1}{10}\,n^{(1)}_2 - n^{(1)} + \beta^{(1)}\chi\,\boostO n^{(0)}\right]
}
using the Thomson collision term, Eq.~\eqref{eq:Thomson_and_Doppler}.
Here, we defined $n_\ell(t, x, \vek{r}, \vgh)=\sum_{m} n_{\ell m}(t, x, \vek{r}) Y_{\ell m}(\vgh)$ using the spherical harmonic coefficients of the photon occupation number, $n_{\ell m}(t, x, \vek{r})$. This definition implies $n^{(0)}\equiv n^{(0)}_0=n^{(0)}_{00} Y_{00}=n^{(0)}_{00}/\sqrt{4\pi}$ for the average spectrum, $n^{(0)}_0=\int n^{(0)}\,\frac{\id \Omega}{4\pi}$, where the solid angle is $\id \Omega=\id^2 \vgh=\id \phi \id \chi$. 

We now assume that the average spectrum is spectrally frozen, $n^{(0)}(x)=\nbb(x)+\Delta n^{(0)}(x)$, where $\Delta n^{(0)}(x)$ is the departure from the blackbody, $\nbb(x)=1/[\expf{x}-1]$ at a temperature $\Tz$. We can then make the simple Ansatz $n^{(1)}\approx \Theta^{(1)}\,\Gspec(x)+ \Sigma^{(1)}\,\boostO \Delta n^{(0)}$.\footnote{For the considered case, one could also directly use $n^{(1)}\approx \Sigma^{(1)}\,\boostO n^{(0)}$, but then the initial condition can not be as clearly separated and an extension to the time-dependent case is not as straightforward.} Assuming that $\boostO \Delta n^{(0)}\neq\Gspec(x)$, this results in two identical photon Boltzmann hierarchies for $\Theta^{(1)}$ and $\Sigma^{(1)}$ that are exactly like the standard equation for the temperature perturbations. 
These can be obtained by simply comparing coefficients of the two independent spectral functions or alternatively by taking the number and energy density moments of the Boltzmann equation (see Appendix~\ref{app:Thomson_hier}).
However, in contrast to the temperature perturbations, at a given wavenumber $k$ we start with the initial condition $\Sigma^{(1)}=0$ such that distortion anisotropies are only sourced later when perturbations in the potentials and velocity field appear \citep{Chluba:2x2}. The transfer functions and related power spectra can be computed directly using any Einstein-Boltzmann code. Cross-correlations of the distortion and temperature fields thus probe Doppler and integrated Sachs-Wolfe (ISW) terms. This can result in a novel noise floor for tests of primordial non-Gaussianity from $\mu/y\times T$ correlations, as we highlight in paper III.

\subsubsection{Preparing for more general spectral evolution}
\label{sec:boosts_of_distortions}
As explained above, in the Thomson limit the problem described by Eq.~\eqref{eq:evol_1_lim_1} can in principle be solved with only two independent variables, $\Theta^{(1)}$ and $\Sigma^{(1)}$ [see Eq.~\eqref{eq:evol_1_lim_1_moments_final} for the Boltzmann equations], and the new spectral shape $\boostO \Delta n^{(0)}$ caused by boosts of the average distortion. For example, if the average spectral distortion was a pure $\mu$-distortion, the anisotropies would have the spectrum of $\boostO M(x)$. We could then observationally search for $G(x)$ and $\boostO M(x)$ and thereby extract the information on the perturbations, $\Theta^{(1)}\propto \zeta$ and $\Sigma^{(1)}\propto \epsilon_\rho \zeta$.

In more general situations, when spectral evolution is also included, using the computational spectral basis we can write $\Delta n^{(0)}\approx\vek{y}^{(0)}\cdot \vek{B}$ and hence
\begin{align}
\label{eq:boost_Dn0}
\boostO \Delta n^{(0)}=\Theta^{(0)}[3\Gspec(x)+\Yspec(x)]+4\sum_{k=1}^{N} y^{(0)}_{k-1}\,Y_{k}(x)+y^{(0)}_{N}\,\boostO Y_{N}(x)+\mu^{(0)} \,\boostO \Mspec(x),
\end{align}
where we used $\boostO G=3G+Y$. The first two groups of terms directly fall back onto the original spectral basis, $\vek{B}$; however, $\boostO Y_{N}(x)=4\Ynspec{N+1}(x)$ and $\boostO \Mspec(x)=[M(x)/G(x)]\boostO G(x)-G(x)/x$ lie outside. As a simple fix, we could add these new spectral shapes to the basis and thereby keep the precision of the average distortion. However, once we consider Compton scattering effects, this approach becomes problematic, requiring a new truncation of the spectral hierarchy.

\begin{figure}
	\centering
	\includegraphics[width=0.95\columnwidth]{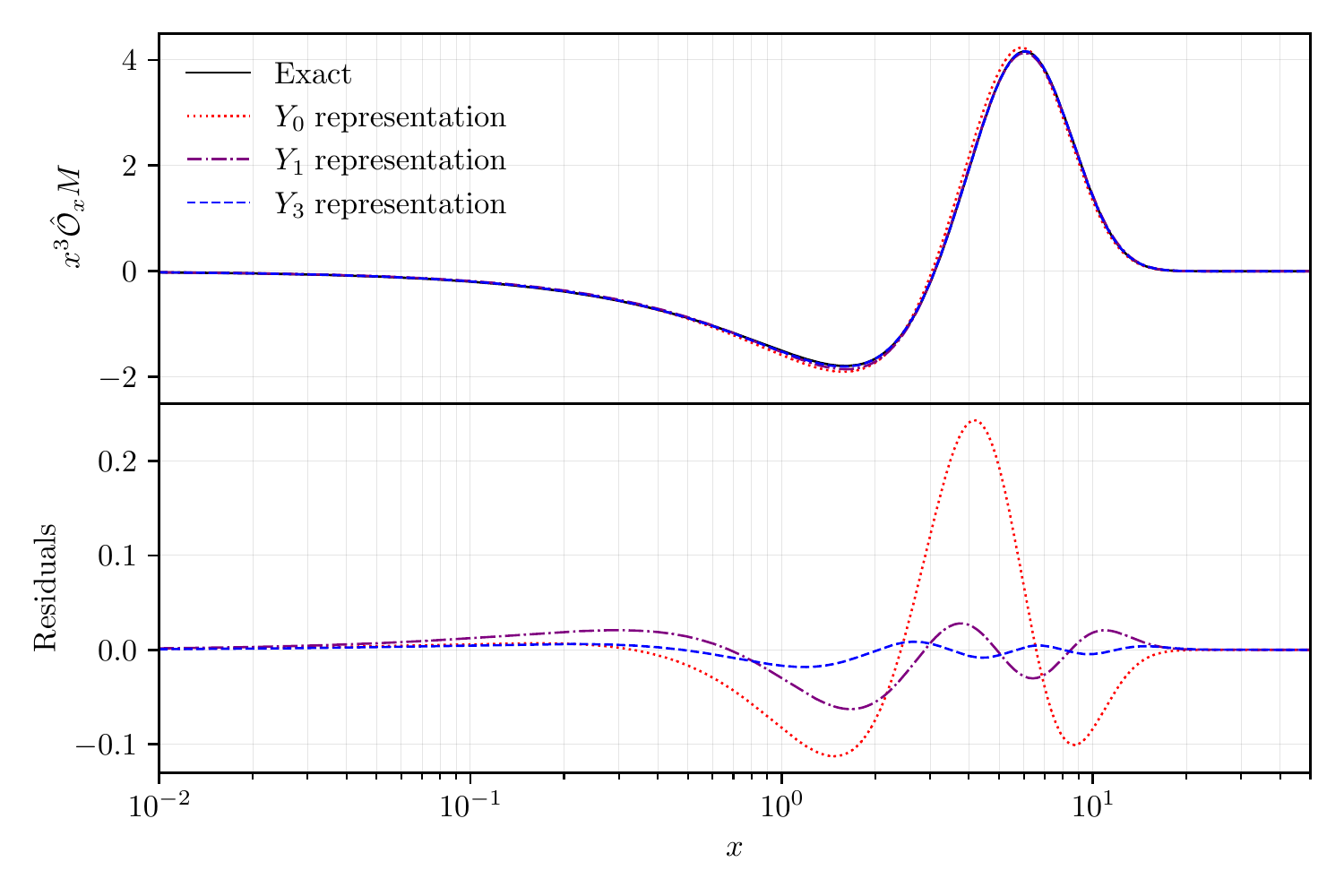}
    \\
	\caption{Representation of $\boostO \Mspec(x)$ using varying elements in the basis. Even the lowest order representation in terms of $\mu$ and $y$ is highly accurate.}
	\label{fig:OM-rep}
\end{figure}
Instead of adding new spectra to the basis, we approximately represented $\boostO Y_{N}(x)$ and $\boostO \Mspec(x)$ within the old spectral basis.\footnote{We tried several alternatives but found none of them to be more beneficial. A more general study is left for the future.}
The projection procedure is explained in detail in paper~I.
Just like for the Kompaneets operator, one has to ensure energy conservation in the calculation; photon number conservation is automatically built in, since $Y_N$ and $M$ do not carry number and hence $\boostO Y_N$ and $\boostO M$ do not either.
It turns out that $\boostO \Mspec(x)$ is extremely well represented as a simple sum of $\mu$ and $y$. Carrying out the projections, we find $\boostO \Mspec(x)\approx 0.3736\,\Yspec(x) +1.9069\,\Mspec(x)$; however, even better precision can be achieved when including additional terms in $Y_k(x)$ (see Fig.~\ref{fig:OM-rep}). As for the Kompaneets operator (see paper I), the largest error is always made when representing $\Ynspec{N+1}(x)$ with only terms up to $Y_{N}(x)$, as part of function space is omitted. However, less and less energy is carried in these contributions such that the precision does not suffer much once $Y_{5}$ or higher are included.

With these comments, schematically we can therefore always write the representations
\begin{align}
\label{eq:OM_OYN_scheme}
\boostO Y_{N}(x)&=4Y_{N+1}(x)\approx \vek{O}_{N}\cdot \vek{B}(x), \qquad 
\boostO \Mspec(x)\approx \vek{O}_{\mu}\cdot \vek{B}(x)
\end{align}
where $\vek{O}_{N}$ and $\vek{O}_{\mu}$ are the corresponding solution vectors in terms of $y, y_1,\ldots,y_N,\mu$ (the projections of $\Gspec(x)$ never matter as $\boostO Y_k$ and $\boostO \Mspec$ do not carry photon number). This means we can express the boosted average spectrum as $\boostO n^{(0)}\approx \vek{b}^{(0)}\cdot \vek{B}$ with 
\bealf{
\label{eq:M_B_def}
\vek{b}^{(0)}=\begin{pmatrix}
1
\\[-0.5mm]
0
\\[-0.5mm]
0
\\[-0.5mm]
0
\\[-0.5mm]
\vdots
\\[-0.5mm]
0
\\[-0.5mm]
0
\\[-0.5mm]
0
\\[-0.5mm]
0
\\[-0.5mm]
\end{pmatrix}\;+\;
M_{\rm B}\,\vek{y}^{(0)}
&=
\begin{pmatrix}
1
\\[-0.5mm]
0
\\[-0.5mm]
0
\\[-0.5mm]
0
\\[-0.5mm]
\vdots
\\[-0.5mm]
0
\\[-0.5mm]
0
\\[-0.5mm]
0
\\[-0.5mm]
0
\\[-0.5mm]
\end{pmatrix}
\;+\;\begin{pmatrix}
     3 &\;\;\;\; 0 &\;\;\;\; 0 &\;\;\;\; 0 &\;\;\;\; \cdots &\;\;\;\; 0 &\;\;\;\; 0 & 0 & 0
     \\[-0.5mm]
     1 &\;\;\;\; 0 &\;\;\;\; 0 &\;\;\;\; 0 &\;\;\;\; \cdots &\;\;\;\; 0 &\;\;\;\; 0 & O_{N,1} & O_{\mu,1}
     \\[-0.5mm]
     0 &\;\;\;\; 4 &\;\;\;\; 0 &\;\;\;\; 0 &\;\;\;\; \cdots &\;\;\;\; 0 &\;\;\;\; 0 & O_{N,2} & O_{\mu,2}
     \\[-0.5mm]
     0 &\;\;\;\; 0 &\;\;\;\; 4 &\;\;\;\; 0 &\;\;\;\; \cdots &\;\;\;\; 0 &\;\;\;\; 0 & O_{N,3} & O_{\mu,3}
     \\[-0.5mm]
     \vdots &\;\;\;\; \vdots &\;\;\;\; \vdots &\;\;\;\; \vdots&\;\;\;\; \vdots &\;\;\;\; 
     \vdots &\;\;\;\; \vdots &\;\;\;\; \vdots
     & \vdots 
     \\[-0.5mm]
     0 &\;\;\;\; 0 &\;\;\;\; 0 &\;\;\;\; 0 &\;\;\;\; \cdots &\;\;\;\; 0 &\;\;\;\; 0 & O_{N,N-2} & O_{\mu,N-2}
     \\[-0.5mm]
     0 &\;\;\;\; 0 &\;\;\;\; 0 &\;\;\;\; 0 &\;\;\;\; \cdots &\;\;\;\; 4 &\;\;\;\; 0 & O_{N,N-1} & O_{\mu,N-1}
     \\[-0.5mm]
     0 &\;\;\;\; 0 &\;\;\;\; 0 &\;\;\;\; 0 &\;\;\;\; \cdots &\;\;\;\; 0 &\;\;\;\; 4 & O_{N,N} & O_{\mu,N}
     \\[-0.5mm]
     0 &\;\;\;\; 0 &\;\;\;\; 0 &\;\;\;\; 0 &\;\;\;\; \cdots &\;\;\;\; 0 &\;\;\;\; 0 & O_{N,N+1} & O_{\mu,N+1}
\end{pmatrix}
\;\;
\begin{pmatrix}
\Theta^{(0)}
\\[-0.5mm]
y^{(0)}_{0}
\\[-0.5mm]
y^{(0)}_{1}
\\[-0.5mm]
y^{(0)}_{2}
\\[-0.5mm]
\vdots
\\[-0.5mm]
y^{(0)}_{N-2}
\\[-0.5mm]
y^{(0)}_{N-1}
\\[-0.5mm]
y^{(0)}_{N}
\\[-0.5mm]
\mu^{(0)}
\\[-0.5mm]
\end{pmatrix},
}
where we implicitly defined the boosting matrix, $M_{\rm B}$,
noting that $\vek{B}$ has $N+3$ entries starting with $G(x)$ for $k=0$ and ending with $M(x)$ for $k=N+2$. Inserting $n^{(1)}=\vek{y}^{(1)}\cdot \vek{B}$ into Eq.~\eqref{eq:evol_1_lim_1} and rearranging terms we can then write the photon Boltzmann equation in the Thomson limit as
\bealf{
\label{eq:evol_1_lim_1_final}
&\frac{\partial \vek{y}^{(1)}}{\partial \eta}+\vgh\cdot \nabla \vek{y}^{(1)}
\approx
-\vek{b}^{(0)}_0
\left(\frac{\partial \Phi^{(1)}}{\partial \eta}+ \vgh\cdot \nabla\Psi^{(1)} \right)
+
\tau'\left[\vek{y}_0^{(1)}+\frac{1}{10}\,\vek{y}_2^{(1)}-
\vek{y}^{(1)}+\beta^{(1)}\chi\,
\vek{b}^{(0)}_0\right],
}
where we now converted to conformal time, $\eta=\int_0^t c\id t'/a$ with $\tau'=\id \tau/\id \eta=\Ne \sigT a$, and also used $\vek{y}_\ell(t,\vek{r},\vgh)=\sum_{m} \vek{y}_{\ell m}(t,\vek{r}) Y_{\ell m}(\vgh)$ as before for $n_\ell$.
We furthermore made explicit that $\vek{b}^{(0)}$ only has a monopolar dependence, $\vek{b}^{(0)}= \vek{b}^{(0)}_0\equiv \int \vek{b}^{(0)}\frac{\id\Omega}{4\pi}$.
For the example above, $\vek{b}^{(0)}_0$ was {\it time-independent}. In more general situations it becomes a function of time, where $\Delta \vek{b}^{(0)}_0=M_{\rm B}\,\vek{y}^{(0)}_0$ provides sources for distortion and temperature anisotropies. However, before we can treat this problem we need to include the effects of spectral evolution into the equations, as we discuss in Sect.~\ref{sec:Comp_effects}. 

\subsubsection{Changing the background temperature}
\label{app:Shift_in_T}
One of the simplest problems that one can already consider with Eq.~\eqref{eq:evol_1_lim_1_final} is a simple time-dependent change of the background temperature. 
This limit describes average injection of $G$, which physically is hard to establish without efficient thermalisation terms \citep{Chluba2014TRR}, but provides some intuition. Using the zeroth order equation, Eq.~\eqref{eq:zeroth_equation}, with $\Delta n^{(0)}=\Theta_0^{(0)}\,G$, and assuming the change in the temperature is created from a source term, $G\,S_T(t)$, we can easily verify that this leaves the spectrum unchanged:
\bealf{
\label{app:zeroth_CS_DC_photon_temp}
G\,\frac{\partial \Theta_0^{(0)}}{\partial t}
&=G\,S_T
+\dot{\tau}\Thz\left[\Theta_{\rm eq}^{(0)}Y+\KompO \Delta n_0^{(0)}\right]
- \dot{\tau}\,\frac{\Lambda(x, \Thz)\,(1-\expf{-x})}{x^3}\Delta n^{(0)}_0+
\dot{\tau}\,\frac{\Lambda(x, \Thz)}{x^2}\,\nbb\,\Theta^{(0)}_{\rm eq}
\nonumber\\[-0.5mm]
&=G\,S_T
+\dot{\tau}\Thz\,\Theta_0^{(0)}\left[Y+\KompO G\right]
-\dot{\tau}\,\frac{\Lambda(x, \Thz)}{x^2}\,\nbb\,\Theta^{(0)}_0
\left[\frac{\Delta n^{(0)}_0}{\Theta^{(0)}_0 G}-1\right]
\equiv G\,S_T.
}
Here, we used $\Theta^{(0)}_{\rm eq}=\Theta_0^{(0)}$ and $(1-\expf{-x})/x=\expf{-x}(\expf{x}-1)/x=\nbb/G$ and assumed that the external source has a spectrum $G$.

\noindent
Knowing the solution of $\bar{\Theta}(t)=\Theta^{(0)}(t)$, from equation~\eqref{eq:evol_1_lim_1_final}, we then find the two hierarchies
\bsub
\label{eq:Temp_evol_with_pert}
\bealf{
\frac{\partial \Theta^{(1)}}{\partial t}+\frac{c\vgh}{a}\cdot \nabla \Theta^{(1)}+
\bar{b}\left(\frac{\partial \Phi^{(1)}}{\partial t}+ \frac{c\vgh}{a}\cdot \nabla\Psi^{(1)}\right)
&\approx \dot\tau\left[\Theta^{(1)}_{0}+\frac{1}{10}\,\Theta^{(1)}_{2} - \Theta^{(1)} + \bar{b}\,\beta^{(1)}\chi\right].
\\[1mm]
\frac{\partial y^{(1)}}{\partial t}+\frac{c\vgh}{a}\cdot \nabla y^{(1)}+
\bar{\Theta}\left(\frac{\partial \Phi^{(1)}}{\partial t}+ \frac{c\vgh}{a}\cdot \nabla\Psi^{(1)}\right)
&\approx \dot\tau\left[y^{(1)}_{0}+\frac{1}{10}\,y^{(1)}_{2} - y^{(1)} + \bar{\Theta}\,\beta^{(1)}\chi\right].
}
\esub
with $\bar{b}(t)=(1+3\bar{\Theta})$. This indicates that in addition to small modifications to the temperature anisotropies, $y$-distortion anisotropies are sourced. The latter are due to the mismatch of the thermal spectrum of the initial anisotropies with respect to the new average blackbody. In the absence of thermalisation effects, this indeed causes distortion anisotropies, as we illustrate now.

To make progress, we first ask the question how a spatially-varying blackbody spectrum changes when the average temperature is varied in a way that departs from the standard $\propto (1+z)$ scaling. For this, we write the expression 
\vspace{-3mm}
\bealf{
\label{app:Ansatz_BB}
n=\frac{1}
{\exp\left(h\nu/[k \bar{T} (1+\Theta)]\right)-1}
\equiv
\nbb\left( x/[(1+\bar{\Theta})(1+\Theta)]\right)
}
where $\bar{T}=\Tz (1+\bar{\Theta})$ and $\Theta$ describes the temperature anisotropies. At zeroth (no spatial terms) and first order in perturbations (only up to terms $\epsilon_\rho \zeta$) this implies
\bsub
\label{app:Ansatz_BB_pert}
\bealf{
n^{(0)}&= \nbb(x) + \bar{\Theta}^{(0)}\,G(x) 
\\[1mm]
n^{(1)}&=  \bar{\Theta}^{(1)}\,G(x)
+\Theta^{(1)}\,[G(x)+\bar{\Theta}^{(0)}\boostO G]
=
[\bar{\Theta}^{(1)}+\Theta^{(1)}(1+3 \bar{\Theta}^{(0)})]\,G(x)
+\bar{\Theta}^{(0)}\Theta^{(1)}\,Y(x),
}
\esub
where we neglected terms $\mathcal{O}(\bar{\Theta}^{(0)})^2$ and $\mathcal{O}(\Theta^{(1)})^2$.
We comment immediately on the consistency of this limit: Terms $\mathcal{O}(\bar{\Theta}^{(0)})^2$ do not add any spatial effects and thus are merely higher order corrections to the average spectrum. Terms relating to $\mathcal{O}(\Theta^{(1)})^2\simeq \mathcal{O}(\zeta)^2$ are second order in the primordial curvature perturbations, which we also do not consider here.

The expressions in Eq.~\eqref{app:Ansatz_BB_pert} seems to suggest that a $y$-type distortion is sourced at first order, even if we started with a blackbody. However, the terms shown above are merely needed to precisely transition from a blackbody at the initial temperature to a new blackbody when $\bar{\Theta}$ varies \citep[e.g.,][]{ Chluba2004, Stebbins2007}. Does this Ansatz solve the evolution equations in Eq.~\eqref{eq:Temp_evol_with_pert}? To answer this question, we first obtain the evolution equation for $\bar{\Theta}^{(1)}$. 
If we take the ensemble average of the full photon Boltzmann equation (to extract the first correction to the `average' evolution equation), we find no source terms, since all terms $\propto \Theta^{(1)}$ and $\bar{\Theta}^{(0)}\,\Theta^{(1)}$ vanish. This means $\bar{\Theta}^{(1)}=0$, as expected. 
This of course assumes that corrections $\propto\bar{\Theta}^2$ are negligible (i.e., small average energy release) since otherwise our `blackbody source' term would have to include a contribution $\propto Y(x)$ to not distort the spectrum (see Appendix~\ref{app:temp_change_term_second}). This statement again emphasises that the appearance of terms $\propto Y(x)$ does not immediately imply a real distortion. 

We can now insert the Ansatz in Eq.~\eqref{app:Ansatz_BB_pert} into the evolution equations in Eq.~\eqref{eq:Temp_evol_with_pert}. For brevity we revert back to the shorthand notation $\bar{\Theta}=\bar{\Theta}^{(0)}$.
Using $\mathcal{D}_t[X]$ from Eq.~\eqref{eq:evol_1_lim_1_operators_a}, we have
\bealf{
\label{app:Ansatz_BB_pert_Dt}
\mathcal{D}_t[n^{(1)}]&=  \mathcal{D}_t[\Theta^{(1)}]\,G(x)
+\mathcal{D}_t[\bar{\Theta}\Theta^{(1)}]\boostO G
=
\mathcal{D}_t[\Theta^{(1)}]\,[G(x)
+\bar{\Theta}\boostO G]
+
\mathcal{D}_t[\bar{\Theta}]
\Theta^{(1)}\boostO G
\nonumber\\[1mm]
&=
\mathcal{D}_t[\Theta^{(1)}]\,\boostO n^{(0)}
+
\Theta^{(1)}\,S_T\,\boostO G,
}
where $\boostO n^{(0)}\equiv G(x)
+\bar{\Theta}\boostO G$ and $\mathcal{D}_t[\bar{\Theta}]=\dot{\bar{\Theta}}=S_T$. Put together with the rest of the Liouville operator [see Eq.~\eqref{eq:evol_1_lim_1_operators} for relevant definitions], we then have
\bealf{
\label{app:Ansatz_BB_pert_L}
\mathcal{L}[n^{(1)}]&= \mathcal{L}[\Theta^{(1)}]\,\boostO n^{(0)}
+
\Theta^{(1)}\,S_T\,\boostO G\equiv \boostO n^{(0)}\mathcal{C}_T[\Theta^{(1)}]+\mathcal{C}^{(1)}_{\rm therm}[n],
}
where we equated with the collision term, formally including both the Thomson terms, $\mathcal{C}_T[\Theta^{(1)}]$, and thermalisation effects, $\mathcal{C}^{(1)}_{\rm therm}[n]$. We also used $n^{(1)}\approx \Theta^{(1)}\,\boostO n^{(0)}$ to factor $\boostO n^{(0)}$ out of $\mathcal{C}^{(1)}_T[n]$.

If we now only consider the Thomson terms and integrate photon number and energy density, with $\int x^2 \boostO n^{(0)}\id x=3N_z(1+3\bar{\Theta})$, $\int x^3 \boostO n^{(0)}\id x=4\rho_z(1+4\bar{\Theta})$, $\int x^2 \boostO G\id x=9 N_z$ and $\int x^3 \boostO G\id x=16\rho_z$, we obtain the two equations
\bealf{
\label{app:Ansatz_BB_pert_moment}
\int x^2 \id x \quad \rightarrow \quad \mathcal{L}[\Theta^{(1)}]
+3
\Theta^{(1)}\,S_T
\approx \mathcal{C}_T[\Theta^{(1)}],
\nonumber
\\
\int x^3 \id x \quad \rightarrow \quad \mathcal{L}[\Theta^{(1)}]
+4
\Theta^{(1)}\,S_T
\approx \mathcal{C}_T[\Theta^{(1)}].
}
Here, we used the number density, $N_z=\frac{8\pi }{c^3}\left(\frac{\kB\Tz}{h}\right)^3\,N_{\nbb}\approx 411.0\,\cm^{-3}\,[T_0(1+z)/2.725\,{\rm K}]^3$, of a blackbody at a temperature $\Tz$ and also neglected higher order terms in $\bar{\Theta}$.
Since both equations have to be fulfilled, this shows that the spectrum cannot be a blackbody anymore without additional thermalisation processes. The reason is that $Y(x)$ does not carry photon number and only contributes to the second equation, leading to an extra source of $\Theta^{(1)}\,S_T$.

Since the only source spectra that are present in the Ansatz are $G$ and $Y$, in the Thomson limit we need an extra independent $y$-parameter to describe the full spectrum. Making the modified Ansatz, 
$n^{(1), \rm m}=n^{(1)}+ y^{(1)}_{\rm d}\,Y(x)$, we then have
\bealf{
\label{app:Ansatz_BB_pert_L_mod}
\mathcal{L}[n^{(1, \rm m)}]&= \left(\mathcal{L}[\Theta^{(1)}]\,(1+3\bar{\Theta})
+3 \Theta^{(1)}\,S_T\right)G 
+\left(\bar{\Theta}\mathcal{L}[\Theta^{(1)}]+\Theta^{(1)} S_T+\mathcal{D}_t[y^{(1)}_{\rm d}]\right) Y 
\nonumber\\
&
\equiv \mathcal{C}_T[\Theta^{(1)}] (1+3\bar{\Theta}) G
+
\left(\bar{\Theta}\,\mathcal{C}_T[\Theta^{(1)}]+\mathcal{C}_T[y^{(1)}_{\rm d}]-\beta^{(1)}\chi\right)Y
+\mathcal{C}^{(1)}_{\rm therm}[n].
}
Here, we explicitly split the terms $\propto G$ and $\propto Y$, since we now wish to keep track of distortion contributions.
By comparing coefficients we can find the modified system
\bsub
\bealf{
\label{app:Ansatz_BB_pert_mod}
\mathcal{L}[\Theta^{(1)}]
+3
\Theta^{(1)}\,S_T
&\approx \mathcal{C}_T[\Theta^{(1)}],
\\
\bar{\Theta}\mathcal{L}[\Theta^{(1)}]
+\Theta^{(1)}\,S_T
+\mathcal{D}_t[y^{(1)}_{\rm d}]
&\approx \bar{\Theta}\,\mathcal{C}_T[\Theta^{(1)}]+\mathcal{C}_T[y^{(1)}_{\rm d}]-\beta^{(1)}\chi,
}
\esub
where we again neglected thermalisation effects. With $\bar{\Theta}\mathcal{L}[\Theta^{(1)}]\approx \bar{\Theta}\,\mathcal{C}_T[\Theta^{(1)}]$, after dropping terms $\simeq \mathcal{O}(\bar{\Theta})^2$, one then has the explicit evolution equations for $\Theta^{(1)}$ and $y^{(1)}_{\rm d}$
\bsub
\label{eq:new_description}
\bealf{
\frac{\partial \Theta^{(1)}}{\partial t}+\frac{c\vgh}{a}\cdot \nabla \Theta^{(1)}+
\frac{\partial \Phi^{(1)}}{\partial t}+ \frac{c\vgh}{a}\cdot \nabla\Psi^{(1)}
&\approx -3\Theta^{(1)}\,S_T+
\dot\tau\left[\Theta^{(1)}_0+\frac{1}{10}\,\Theta^{(1)}_2 - \Theta^{(1)} + \beta^{(1)}\chi\right],
\\
\frac{\partial y^{(1)}_{\rm d}}{\partial t}+\frac{c\vgh}{a}\cdot \nabla y^{(1)}_{\rm d}
&\approx -\Theta^{(1)}\,S_T+
\dot\tau\left[y^{(1)}_{\rm d,0}+\frac{1}{10}\,y^{(1)}_{\rm d,2} - y^{(1)}_{\rm d,0} 
\right].
}
\esub
These equations show that a non-vanishing $y$-parameter is created as perturbations mix through Thomson scattering and free streaming in the presence of average CMB temperature changes. As alluded to above, the reason is that the spectrum of the initial temperature perturbations is not consistent with that of the evolving average blackbody. The correction to the average blackbody sources $\Theta^{(1)}$ and $y^{(1)}_{\rm d}$ perturbations but on average no photons are created in the fluctuating part, implying that the perturbed photon field is not consistent with that of pure blackbody temperature fluctuations. 

This discussion also shows that the Ansatz in Eq.~\eqref{app:Ansatz_BB_pert} can be recast in terms of the effective parameters $\Theta^{(1)}_{\rm eff}=\Theta^{(1)}(1+3\bar{\Theta})$ and $y^{(1)}_{\rm eff}=y^{(1)}_{\rm d}+\bar{\Theta}\Theta^{(1)}$. Indeed, using this redefinition with Eq.~\eqref{eq:new_description} we recover Eq.~\eqref{eq:Temp_evol_with_pert}.
This clearly separates the origin of the distortion anisotropies, identifying temperature raising $y$-contributions, $y^{(1)}_{\rm mix}=\bar{\Theta}\Theta^{(1)}$ from contributions that cannot thermalise, $y^{(1)}_{\rm d}$. These terms can in principle be distinguished from simple higher order corrections $\propto (\Theta^{(1)})^2$ or uncertainty $\propto \Delta T_0$ in the exact average present-day blackbody temperature $\propto \Delta T_0 \Theta^{(1)}$. For the former, this is clear from the fact that the related terms exhibit a different correlation structure. For the latter, the effect would be time independent, which does change the correlation structure. However, a more in depth discussion of these effects is beyond the scope of this work.

We mention that corrections $\propto \bar{\Theta}\,\zeta$ to the potentials and baryon velocity equations in principle also need to be added to the Boltzmann hierarchy. For the velocity terms, we will more explicitly discuss this in Sect.~\ref{sec:beta_evol_eq}. However, here we are not interested in computing the {\it exact} corrections to the temperature power spectra caused by changes of the background temperature, nor will we consider cases when these terms become very important. For computing the physical $y$-parameter caused by changes in the average CMB temperature, we can simply use the old set of equations to solve for $\Theta^{(1)}$, since all corrections $\propto \bar{\Theta}\,\zeta$ recursively enter the evolution equation for $y^{(1)}$ at higher order in $\bar{\Theta}$. In the Thomson limit, we have thus completed the formulation of the problem and demonstrated that real $y$-type distortion anisotropies are created by changing the average temperature. These could potentially be used to test for departures from the standard CMB temperature-redshift relation; however, we leave a more detailed discussion to future work.

\subsubsection{Instantaneous thermalisation}
\label{sec:full_thermalisation}
We can extend our discussion to the case where thermalisation is always instantaneous. At the background level we then have a changing temperature according to $\dot{\bar{\Theta}}=S_T$ relative to $\Tz$ like before. In the anisotropies, photon production and Compton scattering would always ensure that the spectrum of the fluctuating part thermalises under {\it energy conservation}. From the arguments leading up to Eq.~\eqref{app:Ansatz_BB_pert_moment}, we can therefore directly write
\bealf{
\label{eq:new_description_instantaneous}
\frac{\partial \Theta^{(1)}}{\partial t}+\frac{c\vgh}{a}\cdot \nabla \Theta^{(1)}+
\frac{\partial \Phi^{(1)}}{\partial t}+ \frac{c\vgh}{a}\cdot \nabla\Psi^{(1)}
&\approx -4\Theta^{(1)}\,S_T+
\dot\tau\left[\Theta^{(1)}_0+\frac{1}{10}\,\Theta^{(1)}_2 - \Theta^{(1)} + \beta^{(1)}\chi\right],
}
which with the parameter $\Theta^{(1)}_{\rm eff}=\Theta^{(1)}(1+4\bar{\Theta})$ can be cast into the equivalent form
\bealf{
\label{eq:new_description_instantaneous_II}
\frac{\partial \Theta^{(1)}_{\rm eff}}{\partial t}+\frac{c\vgh}{a}\cdot \nabla \Theta^{(1)}_{\rm eff}+
\bar{b}\left(\frac{\partial \Phi^{(1)}}{\partial t}+ \frac{c\vgh}{a}\cdot \nabla\Psi^{(1)}\right)
&\approx
\dot\tau\left[\Theta^{(1)}_{\rm eff, 0}+\frac{1}{10}\,\Theta^{(1)}_{\rm eff, 2} - \Theta^{(1)}_{\rm eff, 0} + \bar{b}\beta^{(1)}\chi\right],
}
with $\bar{b}=(1+4\bar{\Theta})$. This expression indicates that one can simplify the computation by scaling all variables (also $\Phi, \Psi$ and $\beta$) accordingly. However, assuming a general time-dependence of $\bar{\Theta}$, new effects on the CMB power spectra would appear if this equation would be valid throughout the recombination era: the time-dependence of the modes would be modified by the evolution of the background temperature and a time-dependent rescaling of the {\it temperature contrast} would be folded into the shape of the CMB power spectra, an effect that can principally be captured with the above equations. Physically, this of course is an academic example to highlight how a coupling between modes and the background can be formulated. 
Indeed, this effect is an apparent 'super-horizon' effect, which becomes even more obvious when we think about phases before BBN, where multiple phase transitions do increase the average blackbody temperature \citep[e.g.,][]{DodelsonBook, Bauman2005}. Due to quasi-instantaneous thermalisation, this leads to a rescaling of the temperature variables at {\it all scales}, leaving the adiabatic nature of the initial perturbations totally unchanged and merely modifying the initial conditions to account for this modification. Clearly, in this case, there is no tracer of the time-dependent effects that could be observed today, since all modes that witnessed this effect have dissipated away. At the normal CMB scales, we are left with the standard CMB anisotropy evolution (relative to the present-day higher temperature). However, in the primordial gravitational wave background, as an example, the traces of the phase transitions are in principle still visible \citep{Watanabe2006, Kite2022}. 
\subsection{Effect of Compton scattering and photon production}
\label{sec:Comp_effects}
Now that we understand how to account for the effect of boosting and Thomson scattering, we will next include Compton scattering and photon production. These only affect the spectral evolution of the local monopole and can be treated using our ODE representation of the Green's function. We will start by a simplistic treatment (Sect.~\ref{sec:Comp_effects_simplistic}) that just uses a perturbed version of the zeroth order thermalisation terms, Eq.~\eqref{eq:Main_Eq_photon_injection}. A more rigorous derivation (Sect.~\ref{sec:Comp_effects_rig}) shows that a few additional terms appear; however, the overall picture does not change crucially.

\subsubsection{Simplistic inclusion of thermalisation terms}
\label{sec:Comp_effects_simplistic}
Starting from the description in Eq.~\eqref{eq:Main_Eq_photon_injection}, one can obtain a simple version for the thermalisation terms at first order in perturbation theory. Perturbing $\dot \tau$ and including the potential perturbation due to the local inertial frame transformation \citep[e.g., following][]{Senatore2009}, one readily finds
\begin{align}
\label{eq:Main_Eq_photon_injection_perturbed}
&\frac{\partial \vek{y}^{(1)}}{\partial t}\Bigg|_{\rm therm}
\approx \dot{\tau}\Thz \left[M_{\rm K}\,\vek{y}^{(1)}
+\vek{D}^{(1)}\right]
+\dot{\tau}\Thz \left(\delta^{(1)}_{\rm b}+\Psi^{(1)}\right)\left[M_{\rm K}\,\vek{y}^{(0)}
+\vek{D}^{(0)}\right]
+\frac{3}{2}\,\dot{\tau}\Thz \Theta_0^{(1)}\,\vek{D}^{(0)}
+\frac{\dot{\vek{Q}}^{(1)}}{4},
\nonumber
\\
&\vek{D}^{(1)}
=
\left(
     \gamma_T\xc\,\mu^{(1)}, 0, 0, \ldots, 0, -\gamma_N\xc\,\mu^{(1)}
\right)^T,\qquad
\dot{\vek{Q}}^{(1)}
=\left(0, \dot{\mathcal{Q}}^{(1)}, 0, \ldots, 0, 0\right)^T
\\
\nonumber
&
\dot{\mathcal{Q}}^{(1)}=\frac{\dot{Q}^{(1)}_{\rm c}}{\rho_z}
+\Psi^{(1)}\frac{\dot{Q}^{(0)}_{\rm c}}{\rho_z}.
\end{align}
While the equation above has been obtained in a simplistic manner, it actually turns out that even a more rigorous derivation does not change the result that much. 
The first group of terms simply describes the spectral evolution of the first order distortion parameters. The second accounts for corrections with respect to the average evolution from perturbations in the electron density and potentials. Here, we used $\dot{\tau}^{(1)}/\dot{\tau}\approx \delta^{(1)}_{\rm b}$, assuming that we are far from the recombination era, such that perturbed recombination effects \citep[e.g.,][]{Novosyadlyj2006, Senatore2009, Khatri2009} can be omitted. The perturbed heating rate similarly includes perturbations in the first order heating term from collisions in the local inertial frame, $\dot{Q}^{(1)}_{\rm c}$, but also the effect of potentials on the zeroth order term.

However, the term $\frac{3}{2}\,\dot{\tau}\Thz \Theta_0^{(1)}\,\vek{D}^{(0)}$ deserves a bit more explanation. It stems from perturbing the critical frequency, $\xc$, with respect to the local photon temperature. We assumed that only DC is relevant for the emission processes, such that the photon emissivity is $\propto \Thg \xc (\Thg)\simeq \Thg^{3/2}$. Inserting $\Thg= \Thz(1+\Theta_0)$ then yields $[\Thg \xc (\Thg)]^{(1)}\approx \Thz \xc (\Thz) \, (3/2) \, \Theta_0^{(1)}$. This term thus relates to perturbed emission effects and modulates that zeroth order term, $\vek{D}^{(0)}$. However, as we show below the coefficient of this term indeed changes to unity when a more careful account for modifications to the local Compton and DC rate is carried out.

\subsubsection{More rigorous treatment}
\label{sec:Comp_effects_rig}
In this section, we now obtain the thermalisation terms in a more rigorous manner. For this we have to follow a few steps as outlined in Appendix~\ref{app:collision_first} and \ref{app:approximations_kin}. After writing the full photon collision term in Appendix~\ref{app:collision_first}, one has to obtain the electron temperature in the given distorted radiation field. For this we make use of the result of \cite{Senatore2009} and include the effect of Compton scattering (Appendix~\ref{app:Compton_energy_exchange}). While thermalisation terms are relevant, the electron temperature will always be extremely close to the Compton equilibrium temperature, which greatly simplifies the problem and leads to an effective heating term in the photon field as we demonstrate in Appendix~\ref{app:heating_Te}. The main result of that section is Eq.~\eqref{app:first_order_CS} for the Compton terms and heating sources. This expression only neglects one stimulated scattering correction, as explained in that section, but otherwise is consistent at order $\mathcal{O}(\epsilon_\rho \zeta)$. 

Translating Eq.~\eqref{app:first_order_CS} into matrix form using our spectral basis, we find the generalized Kompaneets terms at first order in perturbations
\begin{align}
\label{eq:Main_Eq_photon_injection_perturbed_rigorous}
&\frac{\partial \vek{y}^{(1)}}{\partial t}\Bigg|_{\rm K}
\approx \dot{\tau}\Thz M_{\rm K}\,\vek{y}^{(1)}
+\dot{\tau}\Thz \left(\delta^{(1)}_{\rm b}+\Psi^{(1)}\right)M_{\rm K}\,\vek{y}^{(0)}
+\frac{\dot{\vek{Q}}^{(1)}}{4}
+
\dot{\tau}\Thz\,\Theta_0^{(1)} \left[
M_{\rm D}\,\vek{y}^{(0)}
-\vek{S}^{(0)}
\right],
\nonumber
\\
&\vek{S}^{(0)}
=\left(0, \delta_{\gamma,0}^{(0)}+
4 \Theta_{\rm e}^{(0)},
-4 \Theta_{\rm e}^{(0)}
, \ldots, 0, 0\right)^T,
\end{align}
where $M_{\rm D}=(M_{\rm B} -3 I) M_{\rm B}$ is the {\it Doppler matrix} (see Appendix~\ref{app:Doppler_terms}) and $\Theta_{\rm e}^{(0)}=\Theta_{\rm eq}^{(0)}+\dot{Q}^{(0)}_{\rm c}/[4\dot{\tau}\Thz \rho_z]$ is the average electron temperature. 
We also used  $\delta^{(0)}_{\gamma, 0}=E_{\Delta n_0^{(0)}}/E_{\nbb}=4\Theta_0^{(0)}+4\sum_{n} y_{0, n}^{(0)}+\epsilon_M \mu_0^{(0)}$ for the photon energy density perturbation.
Aside from the last group of terms, all the other are already present in Eq.~\eqref{eq:Main_Eq_photon_injection_perturbed}. As explained in Appendix~\ref{app:heating_Te}, the latter do not add any energy to the system, but merely lead to a more minor change in the spectral distortion evolution. 

In Appendix~\ref{app:photon_production_terms} we carry out a careful derivation of the photon production terms. It turns out that at first order in perturbations, a correction to the DC emissivity causes a modification of the related temperature correction term, changing $\frac{3}{2}\,\dot{\tau}\Thz \Theta_0^{(1)}\,\vek{D}^{(0)}\rightarrow \dot{\tau}\Thz \Theta_0^{(1)}\,\vek{D}^{(0)}$ (see Appendix~\ref{app:photon_production_terms_first}). The cause of this change is the precise balance between Compton and DC scattering terms, as explained there. It is expected that the coefficient can change depending on which approximation for the DC emissivity is actually used. In paper III, we will see that this modification is not expected to be as severe, but additional work may be needed.
Put together we then have the thermalisation terms
\begin{align}
\label{eq:Main_Eq_photon_injection_perturbed_rigor_final}
&\frac{\partial \vek{y}^{(1)}}{\partial t}\Bigg|_{\rm therm}
\approx \dot{\tau}\Thz \left[M_{\rm K}\,\vek{y}^{(1)}
+\vek{D}^{(1)}\right]
+\dot{\tau}\Thz \left(\delta^{(1)}_{\rm b}+\Psi^{(1)}\right)\left[M_{\rm K}\,\vek{y}^{(0)}
+\vek{D}^{(0)}\right]
+\dot{\tau}\Thz \Theta_0^{(1)}\,\vek{D}^{(0)}
+\frac{\dot{\vek{Q}}^{(1)}}{4}
\nonumber 
\\
&\qquad\qquad\qquad\qquad 
+ \dot{\tau}\Thz\,\Theta_0^{(1)} \left[
M_{\rm D}\,\vek{y}^{(0)}
-\vek{S}^{(0)}
\right],
\nonumber
\\
&\vek{D}^{(1)}
=
\left(
     \gamma_T\xc\,\mu^{(1)}, 0, 0, \ldots, 0, -\gamma_N\xc\,\mu^{(1)}
\right)^T,\qquad
\dot{\vek{Q}}^{(1)}
=\left(0, \dot{\mathcal{Q}}^{(1)}, 0, \ldots, 0, 0\right)^T
\nonumber
\\
&\vek{S}^{(0)}
=\left(0, \delta_{\gamma,0}^{(0)}+
4 \Theta_{\rm e}^{(0)},
-4 \Theta_{\rm e}^{(0)}
, \ldots, 0, 0\right)^T,
\\
\nonumber
&
\dot{\mathcal{Q}}^{(1)}=\frac{\dot{Q}^{(1)}_{\rm c}}{\rho_z}
+\Psi^{(1)}\frac{\dot{Q}^{(0)}_{\rm c}}{\rho_z}.
\end{align}
As explained in Appendix~\ref{app:kinKomp}, we neglected kinematic corrections to the perturbed thermalisation terms, as these contribute at order $\simeq \beta\,\dot\tau\,\Thz$ and are therefore smaller than the boosting terms $\propto \beta\,\dot\tau$ in the Thomson contributions. We do not expect this to modify the main conclusions although some details might differ during the $\mu$-era. 

\subsection{Final evolution equation}
We are now in the position to write the full evolution equation in terms of the spectral basis $\vek{y}^{(0)}$ and $\vek{y}^{(1)}$. Together with Eq.~\eqref{eq:M_B_def}, \eqref{eq:evol_1_lim_1_final} and Eq.~\eqref{eq:Main_Eq_photon_injection_perturbed_rigor_final} we find
\bsub
\label{eq:evol_1_final}
\bealf{
\label{eq:evol_1_final_a}
\frac{\partial \vek{y}^{(0)}_0}{\partial \eta}
&=\tau' \Thz\left[M_{\rm K}\,\vek{y}^{(0)}_0+\vek{D}^{(0)}_0\right]+\frac{{\vek{Q}'}^{(0)}}{4},
\\
\label{eq:evol_1_final_Yi_1st_ord}
\frac{\partial \vek{y}^{(1)}}{\partial \eta}+\vgh\cdot \nabla \vek{y}^{(1)}
&=-\vek{b}^{(0)}_0\left(\frac{\partial \Phi^{(1)}}{\partial \eta}+ \vgh\cdot \nabla\Psi^{(1)} \right)
+\tau'\left[\vek{y}_0^{(1)}+\frac{1}{10}\,\vek{y}_2^{(1)}-
\vek{y}^{(1)}+\beta^{(1)}\chi\,\vek{b}^{(0)}_0\right]
+\frac{{\vek{Q}'}^{(1)}}{4}
\\ \nonumber
&
\!\!\!\!\!\!\!\!\!\!\!\!\!\!\!\!
+\tau'\Thz\left\{M_{\rm K}\,\vek{y}^{(1)}_0+\vek{D}^{(1)}_0
+\left[\delta_{\rm b}^{(1)}+\Psi^{(1)}\right]\left(M_{\rm K}\,\vek{y}^{(0)}_0+\vek{D}^{(0)}_0\right)
+\Theta^{(1)}_0\left(\vek{D}^{(0)}_0
+
M_{\rm D}\,\vek{y}^{(0)}
-\vek{S}^{(0)}\right)\right\},
}
\esub
where we made it explicit that the zeroth order solution is only a monopole and converted to conformal time, $\eta$. 
The first line accounts for all terms in the Thomson limit and the effect of anisotropic heating. The last row describes the spectral evolution, with $M_{\rm K}\,\vek{y}^{(1)}_0+\vek{D}^{(1)}_0$ determining the main terms and the other being related to perturbed thermalisation effects. 
%
Before we convert this into a photon distortion parameter hierarchy in Fourier space, let us briefly discuss the expected effect of distortions on the other Einstein-Boltzmann equations.

\subsection{Effect of distortion on the other perturbation equations}
\label{sec:beta_evol_eq}
While we have completed our reformulation of the photon evolution equation in the presence of spectral distortions, we still have to consider how the other perturbation equations might be modified. As an example, let us update the momentum exchange equation with the baryons. For this we have to compute the integral $\frac{1}{2}\,\int x^3 \, \chi C^{(1)}[n] \id x \id \chi$. We shall neglect any small corrections from photon emission and absorption terms. Since the Compton terms all only act on the monopole\footnote{This statement changes if kinematic corrections are taken into account, but the correction is suppressed by a factor of $\Thz\ll 1$ in comparison to the Thomson terms.}, these also do not contribute and one is only left with the Thomson scattering terms. The change of the baryon momentum by scattering with photons at first order in perturbations then is \citep{DodelsonBook}
\bealf{
\label{eq:evol_1_beta}
\frac{\partial \beta^{(1)}}{\partial \eta}\bigg|_{\rm T}
&\approx - \tau'\,\frac{\rho_z}{\rho_{\rm b}} \,\frac{1}{2} \int \frac{x^3 \chi}{E_{\nbb}}
\left[n^{(1)}_0+\frac{1}{10}\,n^{(1)}_2 - n^{(1)} + \beta^{(1)}\,\chi\,\boostO n^{(0)}_0\right] \id x \id \chi
\nonumber\\
&=\frac{\tau'}{\i}\,\frac{\rho_z}{\rho_{\rm b}} \int \frac{x^3\!\id x}{E_{\nbb}} 
\left[\tilde{n}^{(1)}_{1} - \i\,\frac{\beta^{(1)}}{3}\,\boostO \tilde{n}^{(0)}_0\right]\quad\longrightarrow
\nonumber\\
\frac{\partial \tilde{\beta}^{(1)}}{\partial \eta}\bigg|_{\rm T}&=\tau'\,\frac{4\rho_z}{3\rho_{\rm b}} \,\left[
3\left(\tilde{\Theta}^{(1)}_{1}+\sum_{k=0}^N\tilde{y}^{(1)}_{k,1}+\frac{\epsilon_M}{4}\,\tilde{\mu}^{(1)}_1\right)
-\tilde{\beta}^{(1)}\left(1+4\tilde{\Theta}^{(0)}_{0}+4\sum_{k=0}^N\tilde{y}^{(0)}_{k,0}+\epsilon_M\,\tilde{\mu}^{(0)}_0\right)
\right]
}
with $\tilde{\beta}^{(1)}\equiv \i \beta^{(1)}$ and $\tilde{f}_\ell(\eta, x, \vek{r})=\frac{\i^\ell}{2}\int P_\ell(\chi) f(\eta, x, \chi, \vek{r})\id \chi = \i^\ell f_{\ell 0}(\eta, x, \vek{r})/\sqrt{4\pi\,(2\ell+1)}$ based on the Legendre polynomials, $P_\ell(\chi)$. Also, $\rho_z$ and $\rho_{\rm b}$ are the background level quantities (without any distortion effects included).

The presence of distortions modifies the momentum exchange by a small amount. For consistency, these terms should be included in the hierarchies presented below. Here, $\rho_z(1+\delta^{(0)}_{\gamma, 0})$ is the total energy density of the average CMB with
$\delta^{(0)}_{\gamma, 0}=4\tilde{\Theta}^{(0)}_{0}+4\sum_{k=0}^N\tilde{y}^{(0)}_{k,0}+\epsilon_M\,\tilde{\mu}^{(0)}_0$. Similarly, the momentum carried by the photon dipole is modified to $\rho_z\,(\tilde{\Theta}^{(1)}_{1}+\delta^{(1)}_{\gamma, 1})$ with $\delta^{(1)}_{\gamma, 1}=\sum_{k=0}^N\tilde{y}^{(1)}_{k,1}+\frac{\epsilon_M}{4}\,\tilde{\mu}^{(1)}_1$. 
In the tight coupling limit, one then has
\bealf{
\label{eq:evol_1_beta_TC}
\tilde{\beta}^{(1)}_{\rm tc}
&\approx
\frac{3\left(\tilde{\Theta}^{(1)}_{1}+\delta^{(1)}_{\gamma, 1}\right)}
{1+\delta^{(0)}_{\gamma, 0}}
\approx
\frac{3\left(\tilde{\Theta}^{(1)}_{1}+\sum_{k=0}^N\tilde{y}^{(1)}_{k,1}+\frac{\epsilon_M}{4}\,\tilde{\mu}^{(1)}_1\right)}
{1+4\tilde{\Theta}^{(0)}_{0}+4\sum_{k=0}^N\tilde{y}^{(0)}_{k,0}+\epsilon_M\,\tilde{\mu}^{(0)}_0}
\approx 3\tilde{\Theta}^{(1)}_{1}
+\mathcal{O}(\tilde{\Theta}^{(1)}_{1} \Delta n^{(0)}).
}
However, in practice the terms $\delta^{(0)}_{\gamma, 0}$ and $\delta^{(0)}_{\gamma, 1}$ are small corrections to the evolution equations of the standard variables unless the average distortion amplitude becomes close to unity, which is ruled out by \COBEF. 
In a similar way, the equations for the potentials 
and neutrinos as well as temperature polarisation states see negligible effects from the presence of distortions unless a full second order treatment is attempted. For our computation below we shall leave all the first order equations unchanged and only add the new equations describing the distortion anisotropies. Unless anisotropic heating effects are explicitly considered, this means that distortion perturbations are directly driven by the standard perturbations known from the evolution of the CMB temperature with no direct distortion anisotropy sources from the distortions themselves, the latter being $\mathcal{O}(\Delta n^{(0)})^2$.

\subsection{Expression in Fourier space}
To obtain the SD transfer functions we now carry out the final step of going to Fourier space. This allows us to obtain the photon transfer functions using a treatment that is similar to that of the standard perturbation equations \citep{CAMB, CLASSCODE}. 
All the variables appearing above are functions of $X(\eta, \chi, \vek{r})$. Going to Fourier space and using the expansion, $X(\eta, \chi, \vek{r})=\sum_\ell (2\ell+1)\,(-\i)^\ell\tilde{X}_\ell(\eta, \vek{r})\,P_\ell(\chi)$ in terms of the Legendre polynomials, we then have \citep{DodelsonBook, Hu1997}
\bealf{
\label{eq:free-streaming}
\frac{\partial X}{\partial \eta}+\vgh\cdot \nabla X
\qquad\longrightarrow \qquad
\frac{\partial X}{\partial \eta}+\i k \chi X
\qquad\longrightarrow \qquad
\frac{\partial \tilde{X}_\ell}{\partial \eta}+k \,
\left(\frac{\ell+1}{2\ell+1}
\tilde{X}_{\ell+1}-\frac{\ell}{2\ell+1}
\tilde{X}_{\ell-1}\right),
}
where $k$ denotes the wavenumber of the mode. The background evolution equation in Eq.~\eqref{eq:evol_1_final} needs no further modification. Carrying out the Fourier and Legendre transformations of all remaining terms, from Eq.~\eqref{eq:evol_1_final} we then find the final photon hierarchy
\bsub
\label{eq:evol_1_final_hierarchy}
\bealf{
\label{eq:evol_1_final_hierarchy_bg}
\frac{\partial \vek{y}^{(0)}_0}{\partial \eta}
&=\tau'\Thz\left[M_{\rm K}\,\vek{y}^{(0)}_0+\vek{D}^{(0)}_0\right]+\frac{{\vek{Q}'}^{(0)}}{4},
\\[1mm]
\label{eq:evol_1_final_hierarchy_0}
\frac{\partial \tilde{\vek{y}}^{(1)}_0}{\partial \eta}
&=-k\,\tilde{\vek{y}}^{(1)}_1\!-\!
\frac{\partial \tilde{\Phi}^{(1)}}{\partial \eta}
\vek{b}^{(0)}_0
+\frac{{\vek{Q}'}^{(1)}}{4}
\\
\nonumber
&\qquad
+\tau'\Thz\left\{
M_{\rm K}\,\tilde{\vek{y}}^{(1)}_0+\vek{D}^{(1)}_0
+\left[\tilde{\delta}_{\rm b}^{(1)}+\tilde{\Psi}^{(1)}\right]\left(M_{\rm K}\,\vek{y}^{(0)}_0+\vek{D}^{(0)}_0\right)
+\tilde{\Theta}^{(1)}_0\left(\vek{D}^{(0)}_0
+
M_{\rm D}\,\vek{y}^{(0)}
-\vek{S}^{(0)}\right)
\right\},
\\
\frac{\partial \tilde{\vek{y}}^{(1)}_1}{\partial \eta}
&=k \,
\left(\frac{1}{3}
\tilde{\vek{y}}_{0}-\frac{2}{3}
\tilde{\vek{y}}_{2}\right)
+\frac{k}{3}\tilde{\Psi}^{(1)} \,\vek{b}^{(0)}_0
-\tau'\left[
\tilde{\vek{y}}^{(1)}_1-\frac{\tilde{\beta}^{(1)}}{3}\vek{b}^{(0)}_0\right],
\\
\frac{\partial \tilde{\vek{y}}^{(1)}_2}{\partial \eta}
&=
k \,
\left(\frac{2}{5}\tilde{\vek{y}}^{(1)}_1-\frac{3}{5}
\tilde{\vek{y}}^{(1)}_3
\right)
-\frac{9}{10}\,\tau'\,\tilde{\vek{y}}^{(1)}_2,
\\
\frac{\partial \tilde{\vek{y}}^{(1)}_{\ell\geq 3}}{\partial \eta}
&=
k \,
\left(\frac{\ell}{2\ell+1}
\tilde{\vek{y}}_{\ell-1}-
\frac{\ell+1}{2\ell+1}
\tilde{\vek{y}}_{\ell+1}\right)
-\tau'\tilde{\vek{y}}^{(1)}_\ell.
}
\esub
Aside from the new terms in the monopole equations and the generalization of the potential and Doppler sources in terms of the distortion parameters this system is identical to the standard brightness equations \citep{DodelsonBook, Ma1995}. Setting the terms $\tau'\Thz$ and $\vek{Q}'$ to zero and using $\vek{b}^{(0)}_0=(1, 0, \ldots, 0)$ these are identically recovered.\footnote{In comparison to \citep{Hu1997} we use $\Theta^{\rm Hu}_\ell=(2\ell+1)\Theta_\ell$, which also is the definition used in \cite{Ma1995}. For $\Phi$ and $\Psi$, we follow the sign convention of \citep{Hu1997}, which means we have $\Phi^{\rm Ma}=-\Phi$ as defined in \cite{Ma1995}.} 

We note, however, that here we neglected corrections from polarisation terms, which affect the quadrupole equation \citep{Kaiser1983, Weinberg1971, Hu1997}. For the standard temperature terms we can include these as usual, but no attempt to correct polarisation effects for the distortion evolution are made here. This means that the damping scales of the temperature and distortion anisotropies will differ slightly. For the distortions, the Thomson scattering terms will be identical \citep{Pitrou2014}; however, additional effects in the thermalisation terms will have to be studied more carefully, a task that is left to future work.

\subsection{Line of sight integration and power spectra}
\label{eq:tools}
To close our discussion of the theoretical aspects, we briefly present the line of sight approach for obtaining the final signal power spectra. In principle the solutions to the photon hierarchy in Eq.~\eqref{eq:evol_1_final_hierarchy} are enough to compute the power spectra and cross-power spectra of two observables $X$ and $Y$:
\begin{align}
C_\ell^{XY}(\eta)
&=
    \frac{2}{\pi}
    \int k^2 \diff k \, P(k)\, \hat{X}_\ell(\eta, k) \, \hat{Y}_\ell(\eta, k)
\end{align}
once the corresponding transfer functions, in this context indicated by the hat, are obtained. However, this brute force approach becomes numerically challenging, as known from the standard CMB anisotropies \citep{CMBFAST}. To simplify matters, we start by Fourier transforming Eq.~\eqref{eq:evol_1_final_Yi_1st_ord}, which yields
\bsub
\label{eq:evol_Yi_1st_ord_Fourier_I}
\begin{align}
&\frac{\partial \vek{y}^{(1)}}{\partial \eta}
    +\i k \chi \,\vek{y}^{(1)}
    +\tau' \vek{y}^{(1)}
    = \,
    \vek{S}_\text{LOS}, \, 
    \\
&\vek{S}_\text{LOS} \equiv 
    -\left(\frac{\partial \Phi^{(1)}}{\partial \eta}+ \i k \chi\Psi^{(1)} \right)\vek{b}^{(0)}_0
+\tau'\left[\vek{y}_0^{(1)}+\frac{1}{10}\,\vek{y}_2^{(1)}+\beta^{(1)}\chi\,\vek{b}^{(0)}_0\right]
+\frac{{\vek{Q}'}^{(1)}}{4}
\\[1mm] 
\nonumber
&\qquad\qquad
+\tau'\Thz\left\{M_{\rm K}\,\vek{y}^{(1)}_0+\vek{D}^{(1)}_0
+\left[\delta_{\rm b}^{(1)}+\Psi^{(1)}\right]\left(M_{\rm K}\,\vek{y}^{(0)}_0+\vek{D}^{(0)}_0\right)
+\Theta^{(1)}_0\left(\vek{D}^{(0)}_0
+
M_{\rm D}\,\vek{y}^{(0)}
-\vek{S}^{(0)}\right)
\right\},
\end{align}
\esub
where now all variables are functions $f^{(1)}=f^{(1)}(\eta, \chi, k)$. Realizing that the only real differences with respect to the temperature only case are that we now are dealing with a solution vector and a vector for the sources, following the standard steps (see Appendix~\ref{app:LOS}) we can directly write down the solution for $\tilde{\vek{y}}^{(1)}_\ell(\eta_f, k)$ at the final conformal time, $\eta_f$, as 
\begin{align}
\label{eq:formal_sol_Leg_fin}
    \tilde{\vek{y}}^{(1)}_\ell(\eta_f, k) &= \int_0^{\eta_f} \id \eta \,g(\eta)\, \tilde{\mathcal{\vek{S}}}_\ell(\eta, \eta_f, k),
    \\
\tilde{\mathcal{\vek{S}}}_\ell(\eta, \eta_f, k)&=
\left[
 \tilde{\vek{y}}_0^{(1)}+\tilde{\Psi}^{(1)}\vek{b}^{(0)}_0+
 \left(\frac{\partial \tilde{\Psi}^{(1)}}{\partial \eta}
 -\frac{\partial \tilde{\Phi}^{(1)}}{\partial \eta}
 \right)\frac{\vek{b}^{(0)}_0}{\tau'}
 \right]\,j_\ell(k\Delta\eta)
+\tilde{\beta}^{(1)}\vek{b}^{(0)}_0\,j^{(1,0)}_\ell(k\Delta\eta)
+
\frac{\tilde{\vek{y}}_2^{(1)}}{2}\,j^{(2,0)}_\ell(k\Delta\eta)
\nonumber
\\
\nonumber
&\!\!\!\!\!\!\!\!\!\!\!\!\!\!\!\!\!\!\!\!\!\!\!\!\!\!\!\!
+
\!\left\{\Thz\!\left[M_{\rm K}\,\tilde{\vek{y}}^{(1)}_0+\tilde{\vek{D}}^{(1)}_0
+\left[\tilde{\delta}_{\rm b}^{(1)}+\tilde{\Psi}^{(1)}\right]\left(M_{\rm K}\,\vek{y}^{(0)}_0+\vek{D}^{(0)}_0\right)
+\tilde{\Theta}^{(1)}_0 \left(\vek{D}^{(0)}_0
+
M_{\rm D}\,\vek{y}^{(0)}
-\vek{S}^{(0)}\right)
\right]
+\frac{{\vek{Q}'}^{(1)}}{4\tau'}\right\} j_\ell(k\Delta\eta).
\end{align}
Here, we introduced the Thomson visibility function, $g(\eta)=\tau'\,\expf{-\tau_{\rm b}}=\partial_\eta \expf{-\tau_{\rm b}}$ with $\tau_{\rm b}=\tau(\eta_f)-\tau(\eta)$ and the Thomson optical depth $\tau=\int_0^\eta \tau'(\eta') \id \eta'$. We also defined $\Delta \eta=\eta_f-\eta$ for convenience. The functions $j^{(a,b)}_\ell(x)$ are based on the usual spherical Bessel functions $j_\ell(x)$ \citep{Stegun1972}. Concretely we give $j^{(1,0)}_\ell(x)=j'_\ell(x)=\partial_x j_\ell(x)$ and $j^{(2,0)}_\ell(x)=\frac{1}{2}\left[3 j''_\ell(x)+j_\ell(x)\right]$ as in \citep{Hu1997}. 

In comparison to the standard line of sight approach, we have the extra source terms coming firstly from the spectral mixing by Compton scattering and perturbed thermalisation effects ($\propto \Thz$) and secondly from anisotropic heating ($\propto {\vek{Q}'}^{(1)}$). In addition, we have a more general Doppler and potential source vector ($\propto \vek{b}^{(0)}_0$), which in general is time-dependent.

\subsubsection{Reduction to the experimental basis}
\label{sec:basis_rot}
As explained in Sect.~3 of paper I, with an experiment in mind we can reduce the number of spectral parameters by going from the computation basis to the residual distortion representation. From the results for the transfer functions, this can be achieved by applying a matrix $L$ to the solution vector $\vek{y}^{(1)}$. We can then plot the transfer functions for a more limited number of parameters, e.g., $\Theta$, $y$, $r_1$ and $\mu$, without loosing much information. 
In a similar manner we can simplify the computation of power spectra to those of the parameters in the experimental basis. Here is it best to directly apply the rotation to the source vector in the line-of-sight solution, Eq.~\eqref{eq:formal_sol_Leg_fin}. This reduces the computational burden, since the power spectra need to be computed for a reduced number of variables.

We comment that directly applying the projection on the limited observation basis before even computing the transfer function is not an optimal choice as in this case energy conservation is not guaranteed to the same level of precision as with the computation basis. Even if this would further reduce the computational burden, we thus do not recommend such an approach.

\subsection{Basic expectations for the evolution of distortion anisotropies}
\label{sec:understanding}
While it is difficult to anticipate the detailed behavior of the distortion transfer functions by looking at the system in Eq.~\eqref{eq:evol_1_final_hierarchy}, we can already understand some of the most important features. Firstly, without average distortions or anisotropic heating {\it no} distortion anisotropies are generated even in the perturbed universe \citep{Chluba:2x2}. This limit is evident without further explanation, as in this case the system simply becomes identical to the standard temperature anisotropies at first order in perturbation theory. 

Secondly, photon emission and Comptonisation effects $\propto \Thz\simeq \pot{4.6}{-10} (1+z)$ can only be important at $z\gtrsim 10^4$, like for the average distortion evolution. At later times, the spectral evolution is mostly {\it frozen} and only direct distortion sources from boosting and potentials ($\propto \boostO\Delta n^{(0)}$) or anisotropic heating ($\propto \dot{Q}^{(1)} Y$) can create additional distortion anisotropies at significant levels. We note that
if isotropic heating is present
(e.g., by particle decay) it is naturally expected that anisotropic heating occurs unless the heating mechanism is detached from perturbations in the standard cosmic fluid (e.g., independent of the dark matter density or local clock speed). In paper III, we will discuss how distortion anisotropies from decaying or annihilating particles will manifest, and then provide simple constraints on these cases.

Thirdly, the effect of distortions (and thermalisation effects) on the evolution of the standard perturbation variables is negligible unless non-standard sources of temperature perturbations are considered. This means that in many scenarios with distortion anisotropies, the transfer functions of the distortion variables will behave like a driven oscillator, following closely the corresponding temperature variables.
This approximation is possible unless we look at the generation and thermalisation of large distortions during the early phases ($z\gtrsim \pot{5}{6}$), where momentarily one could still imagine average distortions $\mu^{(0)}_0\simeq 10^{-2}$ \citep{Chluba2020large, Acharya2021large} and hence efficient conversion into temperature perturbations of noticeable level. However, we leave a more detailed discussion of this regime to future work.

Moving away from the main time-dependent aspects, let us consider the scale-dependence of the distortion evolution. In the absence of anisotropic heating, distortion anisotropies are only generated once average distortions are present. This means that there will be a difference in the transfer functions depending on when in the cosmic evolution of the mode the injection occurs. Crucial moments are horizon crossing, diffusion damping and free-streaming, all known from the standard CMB temperature anisotropies \citep[e.g.,][]{Hu1995CMBanalytic, DodelsonBook}.

If we pick a mode of wavenumber $k$, we can distinguish three main regimes: 
\begin{itemize}

\item[i)] the average distortion is generated after the mode has fully damped away by Silk-damping ($k\gg \kD$, where $\kD\simeq \pot{4.0}{-6} (1+z)^{3/2}\,{\rm Mpc}^{-1}$ is the Silk-damping scale \citep{Silk1968, Kaiser1983}). In this case, no distortion anisotropies are formed because fluctuations in the plasma are no longer present at the corresponding scale.

\item[ii)] the average distortion is generated when the mode is well within the horizon but at $k\lesssim \kD$. In this regime, no sources of distortions from potential variations arise, and distortion anisotropies are generated solely by Doppler terms and perturbed thermalisation effects while at $z\gtrsim 10^4$.

\item[iii)] the average distortion is generated while the mode is still super-horizon. In this case, both Doppler and potential perturbations affect the distortion fluctuations. We will furthermore be sensitive to perturbed thermalisation effects when considering the evolution at $z\gtrsim 10^4$.

\end{itemize}
We will quantify these general expectations in paper III. We also note that once anisotropic heating is included, the statistical properties of the source of heat could be highly non-Gaussian in which case a more general Boltzmann system may have to be solved. In addition, direct sourcing of distortions in the regime $k\gg \kD$ can still lead to significant distortions at very small scales, e.g., by phase transitions or non-linear baryonic physics.

\vspace{-2mm}
\section{Discussion and conclusions}
\label{sec:discussion}
\vspace{-2mm}
This work provides the main formulation of a new photon Boltzmann hierarchy, Eq.~\eqref{eq:evol_1_final} and Eq.~\eqref{eq:evol_1_final_hierarchy}, that allows us to compute the evolution of distortion anisotropies at first order in perturbation theory (Sect.~\ref{sec:Boltzmann-formulation}). Distortion anisotropy sources from Doppler and potential terms as well as anisotropic heating and perturbed thermalisation are accounted for. We also account for the spectral evolution of the distortion based on an approximate ODE treatment for the thermalisation Green's function, which captures most aspects of the full calculation using an extended spectral basis to describe the residual distortion evolution (see paper I).
We furthermore demonstrate that the line-of-sight approach can be generalized to simplify the computation of the CMB signal power spectra (Sect.~\ref{eq:tools}). We briefly explain how the computation can be simplified by converting to a spectral basis that is optimised for the experiment using a basis rotation as explained in paper I (Sect.~\ref{sec:basis_rot})

Overall, this paper is the second step in a series of works discussing the production and evolution of SD anisotropies generated by various physical mechanisms and how these might be constrained with future CMB spectrometers and imagers, enabling more realististic SD anisotropy forecasts over a wide range of physics. 
The new formulation furthermore is a first step towards a more general and precise treatment of CMB temperature anisotropies at second order in perturbation theory. In the standard approach \citep{Bartolo2006, Pitrou2009, Senatore2009}, an `instantaneous thermalisation' approximation is applied which ensures full energy conservation for the photons but without allowing a rigorous separation of distortion and temperature terms. %
With our new ODE representation of the distortion Green's function, one should be able to overcome this problem, even if additional generalisation will be needed.

We highlight that Doppler boosting and potential driving distortion source terms were omitted in most previous discussions of primordial SD anisotropies, although the importance of these terms was recognized earlier \citep{Chluba:2x2}. These terms are indeed small with respect to the standard temperature perturbation (hence not affecting the their evolution notably); however, for the SD anisotropies they provide the leading order source terms once average distortions are present. Since SDs can be spectrally distinguished, these terms remain relevant, leading to the independent distortion parameter hierarchies we presented here. The resultant distortion anisotropies are expected to be significant as long as the average distortion is at the level that \COBEF allows. A more quantitative discussion will be given in paper III, and subsequent works. Importantly, in paper III we will demonstrate that this effect allows one to derive limits on the average heating rate using existing and future CMB anisotropy measurements.

Finally, the generalised Boltzmann system can still be improved.
We did not consider the effect of polarised distortions nor thermalisation/Compton scattering in the anisotropies, which could further lead modify the result. 
We furthermore neglected kinematic corrections as well as small non-linear Comptonisation terms and Comptonisation effects in the anisotropies.
A brute force treatment of the problem for single modes might reveal additional augmentations to the problem that may require more attention.
It would also be extremely important to formulate the problem in alternative gauges, to check the consistency of the equations. 
We leave all these improvements of our method to the future; however, our main results should not be affected. We therefore conclude that one of the main steps towards quasi-exact computations of anisotropic SDs in the perturbed Universe is taken.

{\small
\vspace{-2mm}
\section*{Acknowledgments}
\vspace{-2mm}
We thank Eiichiro Komatsu, Aditya Rotti and Rashid Sunyaev for stimulating discussion.
In addition we thank Antony Lewis for his questions about apparent super-horizon evolution of distortion modes, which encouraged us to carry out a more complete derivation of the equations.
We furthermore thank Nicola Bartolo, Daniele Bertacca, Colin Hill, Rishi Khatri, Sabino Matarrese, Atsuhisa Ota, Enrico Pajer and Nils Sch\"oneberg for comments on the manuscript.
This work was supported by the ERC Consolidator Grant {\it CMBSPEC} (No.~725456).
TK was also supported by STFC grant ST/T506291/1.
JC was furthermore supported by the Royal Society as a Royal Society University Research Fellow at the University of Manchester, UK (No.~URF/R/191023).
AR acknowledges support by the project "Combining Cosmic Microwave Background and Large Scale Structure data: an Integrated Approach for Addressing Fundamental Questions in Cosmology", funded by the MIUR Progetti di Ricerca di Rilevante Interesse Nazionale (PRIN) Bando 2017 - grant 2017YJYZAH.
}


{\small
\bibliographystyle{JHEP}
\bibliography{bibliography,Lit}
}

\newpage

\appendix

\section{Derivation of the photon collision term at first order}
\label{app:collision_first}
In this section, we obtain the required terms of the collision term at first order in perturbations. For this, we start from Appendix~C3 through C7 of \cite{Chluba:2x2} and also consider some modifications due to the transformation to the local inertial frame, adapting the treatments of \cite{Senatore2009, Pettinari2013PhDT}.
The goal is to include the thermalisation effects and energy exchange for the local monopole and also account for all distortion sources that arise at background level.
In the derivation below, we neglect polarisation terms and non-linear corrections to the distortion evolution, e.g., coming from the Kompaneets term due to stimulated recoil. This means we drop terms $\mathcal{O}(\Delta n_0^2)$, linearising the scattering problem; however, we explain why this seems justified in our treatment of the problem, and also briefly outline how one might be able to account for these effects in the future.

\subsection{Thomson terms}
Starting from the standard physics of CMB temperature anisotropies, the Thomson terms and first order Doppler boosts carry over trivially, leading to \citep[e.g.,][]{Bartolo2006}
\bealf{
\label{eq:Thomson_and_Doppler}
\dot{\tau}^{-1}\,\mathcal{C}^{(1)}[n]\big|_{\rm T}
&=n^{(1)}_0+\frac{1}{10}\,n^{(1)}_2 - n^{(1)} +\beta^{(1)}\,\chi\,\boostO n^{(0)}
}
without further ado. Here, we introduced the direction cosine $\chi=\vgh\cdot\vbetah$ between the velocity and photon direction. We also defined $n_\ell(t, x, \vek{r}, \vgh)=\sum_{m} n_{\ell m}(t, x, \vek{r}) Y_{\ell m}(\vgh)$ using the spherical harmonic coefficients of the photon occupation number, $n_{\ell m}(t, x, \vek{r})$. 
It is important that, in contrast to the usual treatment, we now include the average distortion in $\boostO n^{(0)}=G+\boostO \Delta n^{(0)}$, as this leads to distortion anisotropies, which would not arise otherwise. This is the leading order source term at the level $\propto \dot\tau \beta^{(1)}\,\boostO \Delta n_0^{(0)}$, which becomes noticeable once the mode enters the horizon. Without average distortions, this term vanished and consequently only temperature fluctuations are sourced.

In Eq.~\eqref{eq:Thomson_and_Doppler}, we neglected small Klein-Nishina corrections $\propto \The=\kB\Te/\me c^2$, as these do not change the spectral shape of the photon field but merely modify the Thomson scattering rates of the dipole, quadrupole and octupole \citep{Chluba:2x2, Chluba2014mSZI}. Some of these corrections have been considered in \cite{Haga2018}.

\subsection{Kompaneets terms}
Let us next consider the Comptonisation terms from the scattering of electrons and photon with non-zero energy exchange, which leads to spectral evolution. We shall include these effects only for the local monopole as Thomson terms are not suppressed for $\ell >0$ and hence dominate the evolution there. The standard Kompaneets equation is \cite{Kompa56, Hu1994pert}
\bealf{
\dot{\tau}^{-1}\,\mathcal{C}[n]\big|_{\rm K}&=\The \DiffO n_0 + \Thz \DiffO^* n_0(1+n_0)
= \Delta \The Y + \The \DiffO \Delta n_0 + \Thz \DiffO^* A\,\Delta n_0  + \Thz \DiffO^* \Delta n^2_0
\nonumber\\
\label{eq:Kompaneets}
&=\Delta \The \left[Y + \DiffO \Delta n_0 \right] + \Thz \KompO \Delta n_0 + \Thz \DiffO^* \Delta n^2_0. 
}
where we used $\DiffO=x^{-2}\partial_x x^4 \partial_x$, $\DiffstarO=x^{-2}\partial_x x^4$, $\KompO=\DiffO+\DiffstarO A$ and $A=1+2\nbb$. From the first group of terms we see that the local difference between the electron and photon temperature is the main source of distortions, leading to injection of $Y$ with a source correction from $\DiffO \Delta n_0$.
We have not yet included the effect of transforming to the local inertial frame, which causes an overall factor of $(1+\Psi)$ that usually only becomes important at second order in perturbation theory \citep{Senatore2009, Pettinari2013PhDT}. However, here we are dealing with a non-vanishing zeroth order term, which then required inclusion of this factor already at first order in perturbation once non-zero distortions are present.

Adding this factor and collecting terms, we then have the Compton collision terms at zeroth and first order in perturbations,
\bealf{
\label{eq:Comp_zeroth_order}
\left.\frac{\mathcal{C}^{(0)}[n]}{\dot{\tau}}\right|_{\rm CS}&=\Delta \The^{(0)} Y + \Thz \KompO \Delta n^{(0)}_0+\Delta \The^{(0)} \DiffO \Delta n_0^{(0)}+ \Thz \DiffO^* (\Delta n^{(0)}_0)^2
\approx \Delta \The^{(0)} Y + \Thz \KompO \Delta n^{(0)}_0
\\[1mm]
\left.\frac{\mathcal{C}^{(1)}[n]}{\dot{\tau}}\right|_{\rm CS}&=
\Delta \The^{(1)} Y + \Thz \KompO \Delta n^{(1)}_0+\Delta \The^{(1)} \DiffO \Delta n_0^{(0)}+\Delta \The^{(0)} \DiffO \Delta n_0^{(1)}
+ 2\Thz \DiffO^* \Delta n^{(0)}_0 \Delta n^{(1)}_0
\nonumber\\[-0.5mm]
&\qquad \qquad \qquad +\left[\frac{\dot{\tau}^{(1)}}{\dot{\tau}}+\Psi^{(1)}
\right]\left(\Delta \The^{(0)}\,Y+\Thz\,\KompO\,\Delta n^{(0)}_0 \right)
\nonumber\\[-0.5mm]
\label{eq:Comp_first_order}
&\approx 
\Delta \The^{(1)} Y 
+\Thz \KompO \Delta n^{(1)}_0
+
\left[\frac{\dot{\tau}^{(1)}}{\dot{\tau}}+\Psi^{(1)}
\right]
\left.\frac{\mathcal{C}^{(0)}[n]}{\dot{\tau}}\right|_{\rm CS}
+\Delta \The^{(1)} \DiffO \Delta n_0^{(0)}
+ 4\Theta_0^{(1)}\Delta \The^{(0)} \,Y_1.
}
At zeroth order, we neglect the small correction $\propto \Delta \The^{(0)} \DiffO \Delta n_0^{(0)} + \Thz \DiffO^* (\Delta n^{(0)}_0)^2$, which does not modify the leading order picture significantly, being second order in the average distortion. Also in the standard thermalisation computation with {\tt CosmoTherm} this approximation works to extremely high precision unless large distortions are encountered \citep{Chluba2020large, Acharya2021large}.

At first order, the leading source term is again due to differences in the local electron temperature with respect to $\Tz$. However, since Compton equilibrium is reached quickly, $\Delta \The^{(1)}$ will be comparable to that of the local photon temperature perturbation with corrections from the local distortion and heating rate (see Appendix~\ref{app:heating_Te}). This source term therefore mainly captures the modulation of the local thermal equilibrium by the variation of the ambient blackbody temperature.

The term $\Delta \The^{(0)} \DiffO \Delta n_0^{(1)}$ leads to a correction to the Comptonisation time-scale\footnote{More precisely the Doppler broadening and boosting term, $\propto \DiffO$.} of $\Delta n_0^{(1)}$, with the dominant term given by $\Thz \KompO \Delta n^{(1)}_0$. Since $\Delta \The^{(0)}\simeq \mathcal{O}(\Delta n_0^{(0)})$, here we can use $\Delta n_0^{(1)}\approx \Theta_0^{(1)} G$, such that we obtain the source
\bealf{\DiffO \Delta n_0^{(1)}\approx \Theta_0^{(1)}\,\DiffO \boostO \nbb
\equiv \Theta_0^{(1)}\,\boostO \DiffO \nbb = 4\Theta_0^{(1)}\,Y_1,
} 
as used in the last step of Eq.~\eqref{eq:Comp_first_order}. We can also rewrite the term $\DiffO \Delta n_0^{(0)}=(\boostO-3)\boostO \Delta n_0^{(0)}$ with the method described in the main text to simplify the correction $\Delta \The^{(1)} \DiffO \Delta n_0^{(0)}$, which is related to perturbed scattering terms.

The term $\propto 2\Thz \DiffO^* \Delta n^{(0)}_0 \Delta n^{(1)}_0$ modifies the stimulated Compton scattering rate of $\Delta n_0^{(1)}$. This is dominated by the blackbody part, $\propto \DiffO^* A\, \Delta n^{(1)}_0$, within the Kompaneets operator, so that we neglect this correction. It is clear that this term in principle is of similar order as terms that we indeed keep; however, since it merely modifies the exact timing of leading order scattering terms, we believe that its omission does not change the main conclusions significantly. A full numerical solution will be required to check the error this simplification introduces. For this, we can in principle start by mapping this term back onto the spectral basis just like for the Kompaneets operator, $\KompO$. This will yield a representation in matrix form, creating a rotation of the spectral vector like for $\KompO$; however, we leave an exploration of this effect to the future.

The term $\propto \dot\tau^{(1)}/\dot\tau$ are also due to scattering time-scale modulations but this time from perturbations in the electron density, which we include as $\dot\tau^{(1)}/\dot\tau=\Ne^{(1)}/\Ne\approx \delta^{(1)}_{\rm b}$ using the baryon density perturbation, $\delta^{(1)}_{\rm b}$. We shall neglect perturbed recombination effects here \citep[e.g.,][]{Novosyadlyj2006, Khatri2009, Senatore2009}. These should never become important in the regimes we are interested in, when $\Thz$ is not too small already.
The terms $\propto \Psi^{(1)}$ account for the transformation to the local inertial frame \citep{Senatore2009}, which were omitted in \cite{Chluba:2x2}. These terms lead to perturbed scattering effects once an average distortion is present.

To close the discussion of all the effects from Compton scattering, we mention that additional velocity corrections appear that are related to dipole scattering and kinematic effects. However, to simplify the problem we neglect these terms. From \cite{Chluba:2x2}, it is clear that these terms can only be relevant at high redshifts, where significant energy exchange occurs, as we briefly discuss now. 

\subsubsection{Kinematic corrections to the Kompaneets term}
\label{app:kinKomp}
As already mentioned in the main text, we do not include any energy exchange effects mediated by Compton scattering on the anisotropies in the spectrum. This means that in the rest frame of the moving thermal electron distribution one has
\begin{align}
\label{eq:Main_Eq_rest}
\frac{\partial n'(x', \vgh')}{\partial \ysc'}\Bigg|_{\rm K}
&\approx \frac{\The}{\Thz}\,\DiffOp\,n_0'(x')+ \DiffOp^*\, n_0'(x') [1+n_0'(x')],
\end{align}
where the primes on variables denote that they are evaluated in the moving frame.
After a boost, the average occupation number in the moving frame is $n_0'(x')\approx \int n[x(x',\beta,\chi'),\chi(\beta, \chi')]\,\frac{\id \chi'}{2}
\approx n_0(x')$, where $x(x',\beta,\chi')\approx x'(1+\beta \chi')$ and $\chi(\beta, \chi')= (\chi'+\beta)/(1+\beta \chi')$. Here we neglected possible corrections $\mathcal{O}(\beta n_1)$ from aberration effects on the restframe dipole spectrum to the monopole in the moving frame. Using the invariance of $\DiffOp\equiv \DiffO$ and $\DiffOp^*\approx (1-\beta \chi)\DiffO^*$ together with the transformation of the scattering $y$-parameter we then have
\begin{align}
\label{eq:Main_Eq_lab}
\frac{\partial n(x, \vgh)}{\partial \ysc}
\Bigg|_{\rm K}
&=(1-\beta \chi)
\frac{\partial n'(x', \vgh')}{\partial \ysc'}\Bigg|_{\rm K}
\approx (1-\beta \chi)\left\{\frac{\The}{\Thz}\,\DiffO\,n_0(x')+ (1-\beta \chi)\,\DiffO^*\, n_0(x') [1+n_0(x')]\right\}
\nonumber\\
&\approx \frac{\The}{\Thz}\,\DiffO\,n_0(x)+ \DiffO^*\, n_0(x) [1+n_0(x)]
-\beta \chi
\left\{
\frac{\The}{\Thz}\,\DiffO\,n_0(x)+\DiffO^*\, n_0(x) [1+n_0(x)]\right\}
\nonumber\\
&\qquad 
-\beta \chi \,\DiffO^*\,n_0(x) [1+n_0(x)]
+\beta \chi
\left\{
\frac{\The}{\Thz}\,\DiffO +\DiffO^*[1+2\,n_0(x)]\right\} \mathcal{T}(x)
\end{align}
where in the last step we used $x'\approx x(1-\beta \chi)$ and $n_0(x')\approx n_0(x)+\beta \chi\,\mathcal{T}(x)$ with $\mathcal{T}(x)=\boostO n_0(x)$. This result is consistent with Eq.~(C37) in \cite{Chluba:2x2} after omitting dipole scattering effects. Assuming small departures from the average blackbody spectrum, $\Delta n_0\ll 1$, we can insert $n_0=\nbb+\Delta n_0$ and then linearize in $\Delta n_0$, which with $\mathcal{T}=\Gspec+\boostO \Delta n_0$ yields:
\begin{align}
\label{eq:Main_Eq_lab_lin}
\frac{\partial \Delta n(x, \vgh)}{\partial \ysc}
\Bigg|_{\rm K}
&\approx \frac{\Delta \Te}{\Tz}\,\Yspec(x)
+\KompO\,\Delta n_0
-2\beta \chi
\left\{\frac{\Delta \Te}{\Tz}\,\Yspec(x)
+\KompO\,\Delta n_0\right\}
+\frac{\Te}{\Tz}\,\beta \chi\,\left[\Yspec(x)+\DiffO\,\Delta n_0\right]
\nonumber\\
&\qquad 
+\beta \chi
\left\{
\frac{\Delta \Te}{\Tz}\,\DiffO +\KompO
+2\DiffO^*\,\Delta n_0
\right\} G
+\beta \chi\,\KompO \boostO \Delta n_0
\\ \nonumber
&\approx 
\frac{\Delta \Te}{\Tz}\,\Yspec 
+\KompO\,\Delta n_0
-\beta \chi\left\{\frac{\Delta \Te}{\Tz}\,\left[ \Yspec -4\Ynspec{1}\right]
%
+\left[
\KompO - \KompO \boostO 
+
\DiffO^*\,\left(A-2G
\right)
\right] \Delta n_0\right\}
\end{align}
The first two terms are just the standard Kompaneets terms, while the last group of terms accounts for first order velocity corrections. In computations, the terms in braces would be evaluated at background level and are of order $\beta \chi\,\Thz$ per Thomson event. However, given that from Thomson terms we have a source $\simeq \beta \chi \,\mathcal{T}(x)$, these temperature correction terms are merely a small modification to the main source term (suppressed by a factor of $\Thz\ll 1$). It is therefore well-justified to neglect these contribution unless high precision is needed, as already mentioned above. A more rigorous study of these effects is deemed important for obtaining a self-consistent (gauge-independent) formulation of the problem, in particular once scattering corrections from the anisotropies \citep{Chluba:2x2} are included.

\subsection{Photon number changing processes}
Aside from scattering, which conserves photon number, we also have to treat the conversion of distortions into pure temperature shifts. This thermalisation process is mediated by the combined action of Compton scattering and photon emission processes (DC and BR). Modelling the exact evolution of the photon field when DC and BR are included is difficult; however, it is known that the evolution of the high frequency spectrum ($\nu\gtrsim 1\,{\rm GHz}$) is not affected by these processes once the $\mu$-era ends \citep{Sunyaev1970mu, Hu1993, Chluba2014}. We can therefore obtain an approximate description that simply leads to a {\it redistribution of energy} between the $\mu$ parameter and local photon temperature.
The net photon emission and absorption term takes the explicit form \citep{Hu1993, Chluba2011therm, Chluba2014}
\begin{align}
\label{eq:gen_emissionterm}
\frac{\partial n_0}{\partial \tau}\Bigg|_{\rm em}&=\frac{\Lambda(x, \The,\Thg)\,\expf{-x\,\Thz/\The}}{x^3}\left[1-n_0\left(\expf{x\,\Thz/\The}-1\right)\right]
\end{align}
in the local inertial frame. We introduced the photon emissivity, $\Lambda(x, \The, \Thg)$, which for DC scales as $\Lambda(x, \The, \Thg)\propto \Thg^2$ being driven by the high-frequency blackbody photons \citep{Lightman1981, Danese1982, Ravenni2020DC}. For convenience, we shall use the shorthand notation $\Lambda(x, \Thz)\equiv \Lambda(x, \Thz, \Thz)$ below. We neglect kinematic correction to the emission process and also do not consider the energy exchange corrections from this term. These are expected to be negligible and also require a more careful study of the emission process. Some first steps have been outline in \citep{Chluba:2x2}.

Photon production processes are in equilibrium if the CMB occupation number is given by a blackbody at the electron temperature. Thus, defining the distortion with respect to $\nbb(x \,\Thz/\The)$ would simplify several aspects of the computation; however, the invariance of the spectrum under redshifting is no longer guaranteed, such that we will not use this alternative description.

Perturbing the equation and neglecting terms that are higher order in the average distortions, we can find the emission term at zeroth and first order in perturbations as
\bsub
\begin{align}
\label{eq:em_zeroth}
\frac{\partial n^{(0)}_0}{\partial \tau}\Bigg|_{\rm em}
&\approx -\frac{\Lambda(x, \Thz)(1-\expf{-x})}{x^3}\,\Delta n^{(0)}_0+
\frac{\Lambda(x, \Thz)}{x^2}\,\nbb\Theta^{(0)}_{\rm e}
=
-\frac{\Lambda(x, \Thz)}{x^2}\,\frac{\nbb}{G}\left[\Delta n^{(0)}_0-
\Theta^{(0)}_{\rm e}\,G\right]
\\
\label{eq:em_first}
\frac{\partial n^{(1)}_0}{\partial \tau}\Bigg|_{\rm em}
&\approx 
-\frac{\Lambda(x, \Thz)}{x^2}\,\frac{\nbb}{G}\left[\Delta n^{(1)}_0-
\Theta^{(1)}_{\rm e}\,G\right]
+\Theta^{(1)}_0 \frac{\Lambda(x, \Thz)}{x^2}\left[\Delta n^{(0)}_0\,\expf{-x}
+(G-A+1)\,\Theta^{(0)}_{\rm e} \right]
\nonumber\\
&\qquad -
\left[\delta^{(1)}_{\rm b}+\Psi^{(1)}
\right]\frac{\Lambda(x, \Thz)}{x^2}\,\frac{\nbb}{G}\left[\Delta n^{(0)}_0-
\Theta^{(0)}_{\rm e}\,G\right]
\nonumber\\
&\qquad\qquad -
\Theta^{(1)}_0  
\left[
\frac{\partial \ln \Lambda}{\partial\ln\Thg}\Bigg|_{\Thz}+\frac{\partial \ln \Lambda}{\partial\ln\The}\Bigg|_{\Thz}
\right]\frac{\Lambda(x, \Thz)}{x^2}\,\frac{\nbb}{G}\left[\Delta n^{(0)}_0-
\Theta^{(0)}_{\rm e}\,G\right],
\end{align}
\esub
where we used $(1-\expf{-x})/x=\nbb/G$. For the first order equation, we can see that the temperature derivatives of the emission coefficient modify the thermalisation effect. In the DC era, one has $\partial \ln \Lambda/\partial\ln\The\approx 0$ and $\partial \ln \Lambda/\partial\ln\Thg\approx 2$, where the latter signifies that most of the DC emission is driven by the blackbody part of the CMB spectrum \citep{Lightman1981, Chluba2007, Ravenni2020DC}. Corrections to the picture from BR will be neglected here, but can in principle be added using {\tt BRpack} \citep{BRpack}. However, in the BR era, photon production is already nearly frozen, such that our approximations should not make a significant difference to the final distortion evolution.

\section{Evolution in various limits}
\label{app:limits_evol}

\subsection{Evolution equations in Thomson limit}
\label{app:Thomson_hier}
Assuming that the average distortion is frozen, we can use the Ansatz $n^{(1)}\approx \Theta^{(1)}\,\Gspec(x)+ \Sigma^{(1)}\,\boostO \Delta n^{(0)}$ in Eq.~\eqref{eq:evol_1_lim_1}. We then have
\bsub
\label{eq:Ansatz_ints}
\bealf{
\int x^2 n^{(1)}\id x &\approx N_G\,\Theta^{(1)}+ 3\Sigma^{(1)}\, \int x^2\Delta n^{(0)}\id x\equiv N_G \Theta^{(1)}+3N_{\Delta n^{(0)}}\,\Sigma^{(1)}
\\
\int x^3 n^{(1)}\id x &\approx E_G\,\Theta^{(1)}+ 4\Sigma^{(1)}\,\int x^3 \Delta n^{(0)}\id x
=E_G\Theta^{(1)}+ 4 E_{\Delta n^{(0)}}\,\Sigma^{(1)},
}
\esub
where $N_f=\int x^2 f(x)\id x$ is the number integral of the spectral shape, $f(x)$.
Unless the average CMB temperature was affected by the energy release process, we have $N_{\Delta n^{(0)}}\approx 0$ and otherwise $N_{\Delta n^{(0)}}\approx N_G\,\Theta^{(0)}_0$. Taking the number and energy density moments of the Boltzmann equation, Eq.~\eqref{eq:evol_1_lim_1},
with 
\bsub
\label{eq:evol_1_lim_1_operators}
\bealf{
\label{eq:evol_1_lim_1_operators_a}
&\mathcal{D}_t [X]
=\frac{\partial X}{\partial t}+\frac{c\vgh}{a}\cdot \nabla X
\\
&\mathcal{L}[X, b]
=\frac{\partial X}{\partial t}+\frac{c\vgh}{a}\cdot \nabla X+
b\left(\frac{\partial \Phi^{(1)}}{\partial t}+ \frac{c\vgh}{a}\cdot \nabla\Psi^{(1)}\right)
\\
&\mathcal{C}_{\rm T}[X, b]
=\dot\tau\left[X_0+\frac{1}{10}\,X_2 - X + b\,\beta^{(1)}\chi\right]
}
\esub
and Eq.~\eqref{eq:Ansatz_ints} this then yields
\bsub
\bealf{
\label{eq:evol_1_lim_1_moments}
&\mathcal{L}[\Theta^{(1)}]+3\Theta^{(0)}_0 \mathcal{L}[\Sigma^{(1)}]
\approx
\mathcal{C}_{\rm T}[\Theta^{(1)}]
+3 \Theta^{(0)}_0 \mathcal{C}_{\rm T}[\Sigma^{(1)}]
\\[1mm]
&\mathcal{L}[\Theta^{(1)}]+\frac{4E_{\Delta n^{(0)}}}{E_G} \mathcal{L}[\Sigma^{(1)}]
\approx
\mathcal{C}_{\rm T}[\Theta^{(1)}]
+\frac{4E_{\Delta n^{(0)}}}{E_G}\mathcal{C}_{\rm T}[\Sigma^{(1)}],
}
\esub
where we used the shorthand $\mathcal{L}[X]=\mathcal{L}[X,1]$ and $\mathcal{C}_{\rm T}[X]=\mathcal{C}_{\rm T}[X,1]$.
By taking appropriate sums and differences of these equations we can then find
\bealf{
\label{eq:evol_1_lim_1_moments_final}
&\mathcal{L}[\Theta^{(1)}]
\approx
\mathcal{C}_{\rm T}[\Theta^{(1)}],
\qquad
\mathcal{L}[\Sigma^{(1)}]
\approx
\mathcal{C}_{\rm T}[\Sigma^{(1)}],
}
as would have directly followed by comparing coefficients of the two types of spectra, $G$ and $\boostO \Delta n^{(0)}$. As noted in the main text, $\Theta^{(1)}$ starts with initial conditions from inflation while $\Sigma^{(1)}=0$ until the distortion is in place. The equations above do not allow solving the system while $\Delta n^{(0)}$ is evolving. 

\subsection{Changing the average temperature at second order in $\bar{\Theta}$}
\label{app:temp_change_term_second}
In Sect.~\ref{app:Shift_in_T}, we showed that at first order in $\bar{\Theta}$, a photon source term $\propto G(x)$ leaves the spectrum invariant [confirm Eq.~\eqref{app:zeroth_CS_DC_photon_temp}]. Here we now ask what photon source do we need to obtain a blackbody up to second order in $\bar{\Theta}\ll 1$. Writing the Taylor series for a blackbody as $n\approx \nbb(x)+[\bar{\Theta}+\bar{\Theta}^2]\,G(x)+\frac{1}{2}\bar{\Theta}^2\,Y(x)$ and taking the time derivative (denoted with dot at fixed $x$) we have
\bealf{
\dot{n}&\approx  G(x) [ \dot{\bar{\Theta}}+2\bar{\Theta}\dot{\bar{\Theta}}]+\bar{\Theta}\dot{\bar{\Theta}}\,Y(x)\equiv S_G\,[G(x)+f_Y\,Y(x)].
}
If we demand that the average blackbody temperature should change like $\dot{\bar{\Theta}}=S_T$, then we can trivially write $S_G=(1+2\bar{\Theta})\,S_T$. Similarly, we then have 
$\bar{\Theta}\dot{\bar{\Theta}}\equiv S_G\,f_Y$, which implies $f_Y=\bar{\Theta}/[1+2\bar{\Theta}]$. The overall evolution equation is then given by 
$\dot{n}\approx S_T\,[(1+2\bar{\Theta})G(x)+\bar{\Theta}\,Y(x)]$,
which leaves a blackbody spectrum with changing temperature unchanged at second order in $\bar{\Theta}$. Note that the 'blackbody source' is no longer independent of the solution for the blackbody temperature and requires a $y$-type contribution to shift the maximum of $n$ towards higher frequencies. At second order, one thus cannot create a blackbody with a temperature-independent source.

However, we have an alternative way of describing the same situation. Defining the effective temperature and $y$-parameters as $\bar{\Theta}_{\rm eff}=\bar{\Theta}[1+\bar{\Theta}]$ and $\bar{y}_{\rm eff}=\bar{\Theta}^2/2\approx \bar{\Theta}^2_{\rm eff}/2$, we can similarly determine the required source functions for $G$ and Y. For the source of $G$, we then simply need $S_G\equiv \dot{\bar{\Theta}}_{\rm eff}=(1+2\bar{\Theta})\,S_T=\sqrt{1+4\bar{\Theta}_{\rm eff}}\,S_T\approx (1+2\bar{\Theta}_{\rm eff})\,S_T$. Similarly, we have $S_Y=\dot{\bar{y}}_{\rm eff}=\bar{\Theta}S_T=(\sqrt{1+4\bar{\Theta}_{\rm eff}}-1)S_T/2\approx \bar{\Theta}_{\rm eff}\,S_T$.
Since up to second order in $\bar{\Theta}$ we have $\bar{\Theta}^2/2\approx \bar{\Theta}^2_{\rm eff}/2$, this also directly follows from $S_Y\approx \bar{\Theta}_{\rm eff}\,\dot{\bar{\Theta}}_{\rm eff}\approx \bar{\Theta}_{\rm eff}\,S_T$. We therefore find the same source term as above, but with $\bar{\Theta}\rightarrow \bar{\Theta}_{\rm eff}$. The solution for $\bar{\Theta}_{\rm eff}$ is then 
$\bar{\Theta}_{\rm eff}=[\exp(2\int S_t \id t)-1]/2=[\exp(2\bar{\Theta})-1]/2\approx \bar{\Theta}[1+\bar{\Theta}]$, which as expected shows the equivalence of the two approaches.


\section{Physical approximations for the kinetic equation}
\label{app:approximations_kin}

\subsection{Compton Electron temperature and Compton energy exchange}
\label{app:Compton_energy_exchange}
The energy exchange between electrons (and matter) and photons is controlled by the Compton scattering collision term. Smaller corrections due to BR and DC emission processes can be neglected. From Eq.~\eqref{eq:Comp_zeroth_order}, we can directly write the zeroth order energy exchange term as\footnote{The sign is chosen such that for a blackbody spectrum with $\Tg>\Te$ one has $\Lambda_C^{(0)}>0$ or energy flow to the electrons.}
\bsub
\label{app:Lambda_C_I_C}
\bealf{
\Lambda_C^{(0)}
&=-\frac{8\pi h}{c^3}\left(\frac{\kB\Tz}{h}\right)^4
\int x^3 \dot{\tau} \left[\Delta \The^{(0)} Y + \Thz \KompO \Delta n^{(0)}_0 \right]\id x
\equiv
\kappa
\left[
\Theta^{(0)}_{\rm C}-\Theta^{(0)}_{\rm e}
\right]
\\
\kappa &=4\rho_z \dot{\tau}\,\Thz, 
\qquad
\Theta^{(0)}_{\rm C}
\equiv
\Theta^{(0)}_{\rm eq}
\equiv
\eta_{\Delta n_0^{(0)}} = \frac{\int x^3 \Delta n_0^{(0)} \, w_y \id x}{4E_{\nbb}}
\approx \Theta^{(0)}_0+\sum_k \eta_{Y_k}\,y^{(0)}_{k,0}+\eta_M\,\mu^{(0)}_0,
}
\esub
where $\rho_z=\frac{8\pi h}{c^3}\left(\frac{\kB\Tz}{h}\right)^4\,E_{\nbb}$ is the energy density of a blackbody at a temperature $\Tz$ and $\Theta_{\rm e}=\Delta \Te/\Tz$ denotes the relative temperature difference of $\Te$ and $\Tz$. For convenience we also introduced the background Comptonisation rate, $\kappa$, which will be used frequently below. We furthermore used the Compton integral, $\eta_f$, and the energy integral, $E_f=\int x^3 f(x) \id x$ as defined in paper I.

With Eq.~\eqref{eq:Comp_first_order}, but setting $\Psi^{(1)}=0$ to evaluate in the local inertia frame, we can also directly write the first order Compton energy exchange rate as
\bsub
\label{app:Lambda_C_I_C_first}
\bealf{
\Lambda_C^{(1)}&\approx 
\kappa\left\{
-\Theta^{(1)}_{\rm e}
-\frac{\int x^3 \left[\KompO \Delta n^{(1)}_0+\Theta^{(1)}_{\rm e} \DiffO \Delta n_0^{(0)}
\right]\!\id x}{4 E_{\nbb}}
- 4\Theta_0^{(1)}\Theta^{(0)}_{\rm e}
+
\delta^{(1)}_{\rm b}
\left[
\Theta^{(0)}_{\rm C}-\Theta^{(0)}_{\rm e}
\right]
\right\}
\nonumber\\
&=\kappa\left\{
\left[1+\delta^{(0)}_{\gamma, 0}\right]
\left[\Theta^{(1)}_{\rm C}
-\Theta^{(1)}_{\rm e}\right]
+
\left[\delta^{(1)}_{\rm b}+4\Theta_0^{(1)}
\right]
\left[
\Theta^{(0)}_{\rm C}-\Theta^{(0)}_{\rm e}
\right]
\right\}
\\
\Theta_{\rm C}^{(1)}&=\frac{1}{1+\delta^{(0)}_{\gamma, 0}}
\left[
\frac{\int x^3 \Delta n_0^{(1)} \, w_y \id x}{4E_{\nbb}}-
4\Theta_0^{(1)}\,\Theta_{\rm eq}^{(0)}\right]
\approx \Theta_{\rm eq}^{(1)}-\Theta_0^{(1)}\,\delta^{(0)}_{\gamma, 0}
-
4\Theta_0^{(1)}\,\Theta_{\rm eq}^{(0)}.
}
\esub
with $\delta^{(0)}_{\gamma, 0}=E_{\Delta n_0^{(0)}}/E_{\nbb}=4\Theta_0^{(0)}+4\sum_{n} y_{0, n}^{(0)}+\epsilon_M \mu_0^{(0)}$, which accounts for a small correction to the energy density of the zeroth order photon field. We will see that this term ensures energy conservation with respect to the scattering correction, $\Delta \The^{(1)} \DiffO \Delta n_0^{(0)}$ in Eq.~\eqref{eq:Comp_first_order}. 

We comment that the Compton equilibrium temperature corrections, $\Theta_{\rm C}^{(0)}$ and $\Theta_{\rm C}^{(1)}$, could also have been obtained directly using the well-known expression \cite{Levich1970, Sazonov2001}
\bsub
\label{app:Theta_C}
\bealf{
\Theta_{\rm C}
&= \frac{\int x^4 n(1+n)\id x}{4\int x^3 n\id x}-1=\frac{\int x^3 \Delta n_0 \, w_y \id x + \int x^4 \Delta n_0^2 \id x}{4E_{\nbb}+4\int x^3 \Delta n_0 \id x}
}
\esub
both at zeroth and first order in perturbation when non-linear distortion terms ($\propto \Delta n_0^2$) are neglected. The direct path using the Compton collision term highlights the consistency of the result.

\subsection{Electron temperature equation and effective photon heating rate}
\label{app:heating_Te}
In this section, we consider solutions to the local electron temperature including perturbations in the medium. Due to the presence of Compton scattering, the electron temperature is always pushed extremely close to the Compton equilibrium temperature, with corrections from the heating terms that appear in the electron temperature equation. This leads to an effective heating term in the photon equation that is obtained here for conditions in the pre-recombination era. 

To obtain the correct terms, we adapt the discussion of \cite{Senatore2009}. Denoting time-derivatives in this context with `dot', during the pre-recombination era and consistent up to first order in perturbations one has\footnote{We use $\Psi=\Psi^{\rm S}$ and $\Phi=-\Phi^{\rm S}$.}
\bealf{
C_V\left[
\dot{T}_{\rm e}+2H\,T_{\rm e}
+2\left(\frac{\beta}{3a}+\dot{\Phi}\right)T_{\rm e}
\right]
&=(1+\Psi)\left[\Lambda_C+\dot{Q}_{\rm c}\right]
}
for the electron temperature. Here, $C_V=\frac{3}{2}k_{\rm B}\,N_{\rm tot}$ is the non-relativistic heat capacity of the baryons, with $N_{\rm tot}=N_{\rm H}+N_{\rm He}+N_{\rm e}$ being the total number density of baryonic particles. 

Armed with these ingredients we can now write down the evolution equation for the electron temperature at background and perturbed level. This yields
\bealf{
\nonumber
C^{(0)}_V\left[
\dot{T}^{(0)}_{\rm e}+2 H T^{(0)}_{\rm e}
\right]
&=\Lambda_C^{(0)}+\dot{Q}^{(0)}_{\rm c},
\\
\nonumber
C^{(1)}_V\left[
\dot{T}^{(0)}_{\rm e}+2 H T^{(0)}_{\rm e}
\right]
+
C^{(0)}_V\left[
\dot{T}^{(1)}_{\rm e}+2 H T^{(1)}_{\rm e}
+
2\left(\frac{\beta^{(1)}}{3a}+\dot{\Phi}^{(1)}\right) T_{\rm e}^{(0)}
\right]
&=\Psi^{(1)}\left[\Lambda_C^{(0)}+\dot{Q}^{(0)}_{\rm c}\right]+\Lambda_C^{(1)}+\dot{Q}^{(1)}_{\rm c}.
}
Assuming the plasma to be fully ionised one finds $C^{(1)}_V=C^{(0)}_V\,\delta^{(1)}_{\rm b}$. 

The background level equation is well-known in connection with the standard recombination and thermalisation problems \citep[e.g.,][]{Seager2000, Hu1993}. In the pre-recombination era, $\Te$ will follow a sequence of quasi-stationary states due to rapid Compton interactions, pushing $\Te^{(0)}\approx \Tz$. Making the Ansatz $\Te^{(0)}= \Tz + \Delta \Te^{(0)}$ and setting $\Delta \dot{\Te}^{(0)}\approx 0$, we find
\bealf{
\nonumber
&C^{(0)}_V\left[-H \Tz+2 H \Tz\left(1+\Theta_{\rm e}^{(0)}\right)
\right]
\approx \kappa\left[
\Theta_{\rm eq}^{(0)}-\Theta_{\rm e}^{(0)}
\right]
+\dot{Q}^{(0)}_{\rm c}.
\nonumber
\\[1mm]
\label{app:Te_sol_zeroth}
\rightarrow
&\quad
\Theta_{\rm e}^{(0)}=\frac{\Delta \Te^{(0)}}{\Tz}
\approx
\frac{\kappa
\Theta_{\rm eq}^{(0)}
+\dot{Q}^{(0)}_{\rm c} - H C^{(0)}_V  \Tz}{\kappa+2H\, C^{(0)}_V}
\approx \Theta_{\rm eq}^{(0)}+\frac{\dot{Q}^{(0)}_{\rm c}}{\kappa},
}
where in the second step we neglected the correction from the adiabatic cooling effect, which leads to a tiny average distortion $\mu_{\rm cool}\simeq -\pot{3}{-9}$ \cite{Chluba2005, Chluba2011therm, Khatri2011BE}. We could have directly obtained the solution by setting $\Lambda_C^{(0)}+\dot{Q}^{(0)}_{\rm c}\equiv \kappa\left[
\Theta_{\rm eq}^{(0)}-\Theta_{\rm e}^{(0)}
\right]
+\dot{Q}^{(0)}_{\rm c}\approx 0$, a fact that we will use when computing the perturbed temperature solution.
Starting with Eq.~\eqref{app:Te_sol_zeroth} for the electron temperature, we then obtain 
\bealf{
\label{app:zeroth_CS}
\frac{\partial \Delta n_0^{(0)}}{\partial t}\Bigg|_{\rm K}
&=
\dot{\tau}\Thz\left[\Theta_{\rm e}^{(0)}Y +\KompO \Delta n_0^{(0)}\right]
=
\dot{\tau}\Thz\left[\Theta_{\rm eq}^{(0)}Y+\KompO \Delta n_0^{(0)}\right]+\frac{\dot{Q}^{(0)}_{\rm c}}{4\rho_z}\,Y
}
for the Kompaneets terms. 
The combination $\Theta_{\rm eq}^{(0)}Y+\KompO \Delta n_0^{(0)}$ can be replaced with the Kompaneets matrix description $\rightarrow M_K\,\vek{y}^{(0)}$.
Here, $\Theta_{\rm eq}^{(0)}\approx \Theta^{(0)}_0+\sum_k \eta_{Y_k}\,y^{(0)}_{k,0}+\eta_M\,\mu^{(0)}_0$ once $\Delta n_0^2$ terms are neglected. 
From this we can also identify the relative effective heating rate
\bealf{
\dot{\mathcal{Q}}^{(0)}\equiv 
\frac{\dot{Q}^{(0)}_{\rm c}}{\rho_z},
}
which then leads to the formulation for the background distortion evolution given in the main text. 

We note that when exact Compton equilibrium is no longer reached, we can directly solve the zeroth order temperature equation given above to compute the main distortion source term in Eq.~\eqref{app:zeroth_CS}. In this regime, we can neglect the terms $\dot{\tau} \Thz$ and simply find direct sources of $y$-distortions without any spectral evolution. This only is expected to become important at $z\lesssim 200$, which is a regime we do not consider at this point, such that the quasi-stationary solution is valid.

To obtain the perturbed temperature solution, we take the anticipated short-cut assuming quasi-stationary conditions (or more accurately, $C_V H/\kappa\ll 1$). This then yields
\bealf{
0&\approx \Psi^{(1)}\left[\Lambda_C^{(0)}+\dot{Q}^{(0)}_{\rm c}\right]+\Lambda_C^{(1)}+\dot{Q}^{(1)}_{\rm c}.
}
This expression neglects the perturbed corrections to the adiabatic cooling process, which are sub-dominant for our purposes, but should lead to a minimal distortion anisotropy in $\Lambda$CDM. In our limit, $\Lambda_C^{(0)}\approx\kappa\left[\Theta_{\rm eq}^{(0)}-\Theta_{\rm e}^{(0)}\right]\approx -\dot{Q}^{(0)}_{\rm c}$ and hence $\Lambda_C^{(0)}+\dot{Q}^{(0)}_{\rm c}\approx 0$.
With Eq.~\eqref{app:Lambda_C_I_C_first}, this then implies
\bealf{
\nonumber
0&\approx 
\kappa\left\{
\left[1+\delta^{(0)}_{\gamma, 0}\right]
\left[\Theta^{(1)}_{\rm C}
-\Theta^{(1)}_{\rm e}\right]
+
\left[\delta^{(1)}_{\rm b}+4\Theta_0^{(1)}
\right]
\left[
\Theta^{(0)}_{\rm eq}-\Theta^{(0)}_{\rm e}
\right]
\right\}
+\dot{Q}^{(1)}_{\rm c}
\\
&\approx 
\kappa
\left[1+\delta^{(0)}_{\gamma, 0}\right]
\left[\Theta^{(1)}_{\rm C}
-\Theta^{(1)}_{\rm e}\right]
-
\left[\delta^{(1)}_{\rm b}+4\Theta_0^{(1)}
\right]
\dot{Q}^{(0)}_{\rm c}
+\dot{Q}^{(1)}_{\rm c}
}
By solving for $\Theta_{\rm e}^{(1)}$ and neglecting higher order distortion terms, we then finally find
\bealf{
\label{app:Te_sol_first}
\Theta_{\rm e}^{(1)}\approx 
\Theta_{\rm C}^{(1)} 
+
\frac{\dot{Q}^{(1)}_{\rm c}}{\kappa}
-
\left[\delta^{(1)}_{\rm b}+4\Theta_0^{(1)}\right]
\frac{\dot{Q}^{(0)}_{\rm c}}{\kappa}
\approx \frac{\Theta_{\rm eq}^{(1)}}{1+\delta^{(0)}_{\gamma, 0}}
+\frac{\dot{Q}^{(1)}_{\rm c}}{\kappa}
-
4\Theta_0^{(1)} \Theta_{\rm eq}^{(0)}
-
\left[\delta^{(1)}_{\rm b}+4\Theta_0^{(1)}\right]
\frac{\dot{Q}^{(0)}_{\rm c}}{\kappa}
,}
where the term $\propto 4\Theta_0^{(1)}$ accounts for variations of the local photon heat capacity.
Putting everything together, from Eq.~\eqref{eq:Comp_first_order} we find
\bealf{
\frac{\partial \Delta n_0^{(1)}}{\partial t}\Bigg|_{\rm K}
&\approx 
\dot{\tau}\Thz\left[\Theta_{\rm e}^{(1)}Y +\KompO \Delta n_0^{(1)}+\Theta_{\rm e}^{(1)} \DiffO \Delta n_0^{(0)}
+ 4\Theta_0^{(1)} \Theta_{\rm e}^{(0)} \,Y_1
\right]
+\dot{\tau}\Thz\left[\delta_{\rm b}^{(1)}+\Psi^{(1)}\right]\left[\Theta_{\rm e}^{(0)}Y +\KompO \Delta n_0^{(0)}\right]
\nonumber 
\\
&\approx
\dot{\tau}\Thz\left[
(\Theta_{\rm eq}^{(1)}-\Theta_0^{(1)}\delta^{(0)}_{\gamma, 0}) Y +\KompO \Delta n_0^{(1)}
+ 4\Theta_0^{(1)} \Theta_{\rm eq}^{(0)} \,(Y_1-Y)\right]
+
\dot{\tau}\Thz
\left[\delta_{\rm b}^{(1)}+\Psi^{(1)}\right]\left[\Theta_{\rm eq}^{(0)}Y +\KompO \Delta n_0^{(0)}\right]
\nonumber
\\[-1mm]
&\qquad\qquad
+\dot{\tau}\Thz \Theta_0^{(1)} \DiffO \Delta n_0^{(0)}
+
\left[\frac{\dot{Q}^{(1)}_{\rm c}}{\rho_z}
+\Psi^{(1)}\,
\frac{\dot{Q}^{(0)}_{\rm c}}{\rho_z}
\right] \frac{Y}{4}
+ \Theta_0^{(1)} \frac{\dot{Q}^{(0)}_{\rm c}}{\rho_z}\,(Y_1-Y)
\nonumber
\\[-1mm]
&\approx
\dot{\tau}\Thz\left[
\Theta_{\rm eq}^{(1)}\,Y +\KompO \Delta n_0^{(1)}
\right]
+
\dot{\tau}\Thz
\left[\delta_{\rm b}^{(1)}+\Psi^{(1)}\right]\left[\Theta_{\rm eq}^{(0)}Y +\KompO \Delta n_0^{(0)}\right]
+
\left[\frac{\dot{Q}^{(1)}_{\rm c}}{\rho_z}
+\Psi^{(1)}\,
\frac{\dot{Q}^{(0)}_{\rm c}}{\rho_z}
\right] \frac{Y}{4}
\nonumber
\\[-1mm]
\label{app:first_order_CS}
&\qquad\qquad
+\dot{\tau}\Thz \Theta_0^{(1)} \left(\DiffO \Delta n_0^{(0)}-\delta^{(0)}_{\gamma, 0}\,Y\right)
+ 4\dot{\tau}\Thz \,\Theta_0^{(1)} \left[\Theta_{\rm eq}^{(0)}+\frac{\dot{Q}^{(0)}_{\rm c}}{\kappa} \right]\,(Y_1-Y)
}
for the Kompaneets terms at first order. We can identify  $\Theta_{\rm eq}^{(1)}Y+\KompO \Delta n_0^{(1)}=M_K\,\vek{y}^{(1)}$ like for the zeroth order.
We note that in the second line we used the equilibrium temperature differences to isolate the effect of external heating terms. We also set $\Theta_{\rm e}^{(1)} \DiffO \Delta n_0^{(0)}\approx \Theta_0^{(1)} \DiffO \Delta n_0^{(0)}$, given that the other corrections to $\Theta_{\rm e}^{(1)}$ are higher order in the distortion. 

It is important to highlight that the terms in the last line of Eq.~\eqref{app:first_order_CS} do not add any extra energy to the system, but merely redistribute the energy spectrally, leading to a correction to the perturbed scattering effect. For the terms proportional to $\propto Y_1-Y$ this is trivially seen when computing $\int x^3 Y \id x=\int x^3 Y_1 \id x=4 E_{\nbb}$. For the other term we have $\int x^3 \DiffO \Delta n_0^{(0)} \id x=-\int x^4 \partial_x  \Delta n_0^{(0)} \id x=4 \int x^3 \Delta n_0^{(0)} \id x = 4 E_{\nbb} \,\delta^{(0)}_{\gamma, 0}$. With $\int x^3 Y \id x=4 E_{\nbb}$ this identically chancels the remaining term in the last line and also highlights how the term $\propto \delta^{(0)}_{\gamma, 0}$ is crucial for conserving energy. 

The only terms in Eq.~\eqref{app:first_order_CS} leading to addition of energy are those $\propto \dot{Q}^{(1)}_{\rm c}/\rho_z
+\Psi^{(1)}\,
\dot{Q}^{(0)}_{\rm c}/\rho_z$. We thus identify them as 
\vspace{-4mm}
\bealf{
\label{app:pert_heating}
\dot{\mathcal{Q}}^{(1)}\equiv
\frac{\dot{Q}^{(1)}_{\rm c}}{\rho_z}
+ \Psi^{(1)}
\frac{\dot{Q}^{(0)}_{\rm c}}{\rho_z},
}
which is used in the main text to formulate the first order heating rate. We highlight that the terms $\dot{Q}^{(0)}_{\rm c}$ and $\dot{Q}^{(1)}_{\rm c}$ simply follow from the collision terms in the local inertial frame. One would have naturally guessed this perturbed heating term when thinking about $\dot{\mathcal{Q}}\approx (1+\Psi)\dot{Q}_{\rm c}/\rho_z$.

\vspace{-1mm}
\subsubsection{Adding extra Doppler terms to the system}
\label{app:Doppler_terms}
Since the terms in the last line of Eq.~\eqref{app:first_order_CS} do not add any extra energy (or photons) to the system, we can in principle neglect them in our computation, without changing the consistency of the system. However, it is fairly easy to include them once $\DiffO \Delta n_0^{(0)}$ is evaluated. Since they are expected to change the exact distortion evolution across the residual era, this could provide extra time-dependent information.
Inserting our distortion representation, with $\DiffO=-3\boostO+\boostO^2$ we have
\bealf{
\DiffO G&=4Y_1, \qquad \DiffO Y_n=12 Y_{n+1}+16\,Y_{n+2}, \qquad
\DiffO \Delta n_0^{(0)}=4\Theta_0^{(0)}\,Y_1+
\DiffO \Delta n_{0,d}^{(0)}
\nonumber
\\
\DiffO \Delta n_{0,\rm d}^{(0)}&=
4\sum_{n=0}^{N-2} (3 Y_{n+1}+4Y_{n+2})\,y_{0,n}^{(0)}
+\DiffO Y_{N-1} y_{0,N-1}^{(0)}+\DiffO Y_{N}\, y_{0,N}^{(0)}+
\DiffO M \,\mu_{0}^{(0)},
}
where we separated the distortion only terms (i.e., those without $G$). We can in principle again use our projection method to obtain descriptions of $\DiffO Y_{N-1}$, $\DiffO Y_{N}$ and $\DiffO M$ in our computation basis, but it is much easier to just apply the boost matrix, $M_{\rm B}$, defined in Eq.~\eqref{eq:M_B_def} for this purpose. It turns out that we then have
\bealf{
\DiffO \Delta n_0^{(0)}
=(M_{\rm B} -3 I) M_{\rm B} \,\vek{y}^{(0)}=M_{\rm D} \,\vek{y}^{(0)}
}
where $I$ is the identity matrix. One can precompute the {\it Doppler matrix} $M_{\rm D}=(M_{\rm B} -3 I) M_{\rm B}$ for numerical applications to ease the computations.

\vspace{-1mm}
\subsubsection{Compton equilibrium spectrum}
An important property of the Compton collision term operator is the associated equilibrium spectrum, which leads to a stationary spectrum under repeated scatterings. At zeroth order in perturbations, we trivially have $\Delta n_{0,\rm eq}^{(0)}=G \,\Theta_0^{(0)}+ M\,\mu_0^{(0)}$, as can be verified by using $\KompO n_{0,\rm eq}^{(0)}=-Y\,(\Theta_0^{(0)}+\eta_M\,\mu_0^{(0)})$ and $\Theta_{\rm eq}^{(0)}=\Theta_0^{(0)}+\eta_M\,\mu_0^{(0)}$, which leads to cancellation of the Kompaneets term at zeroth order.

How does this work for the first order Comptonisation terms? If we assume no average distortion or temperature shift and no heating, then from Eq.~\eqref{app:first_order_CS} we naturally have 
\bealf{
\frac{\partial \Delta n_0^{(1)}}{\partial t}\Bigg|_{\rm K}
&\approx
\dot{\tau}\Thz\left[
\Theta_{\rm eq}^{(1)}\,Y +\KompO \Delta n_0^{(1)}
\right],
}
which again has the general solution $\Delta n_{0,\rm eq}^{(1)}=G \,\Theta_0^{(1)}+ M\,\mu_0^{(1)}$ and $\Theta_{\rm eq}^{(1)}=\Theta_0^{(1)}+\eta_M\,\mu_0^{(1)}$. Due to the absence of any source terms for $M$ we would even expect $\mu_0^{(1)}=0$. Comptonisation would therefore not modify the spectrum of temperature perturbations, as naturally expected.

Let us now assume that the average spectrum has a slightly higher temperature than $\Tz$, i.e., $\Delta n_{0,\rm eq}^{(0)}=G \,\Theta_0^{(0)}$. In the absence of external heating, from Eq.~\eqref{app:first_order_CS} we find
\bealf{
\frac{\partial \Delta n_0^{(1)}}{\partial t}\Bigg|_{\rm K}
&\approx 
\dot{\tau}\Thz\left[
\Theta_{\rm eq}^{(1)}\,Y +\KompO \Delta n_0^{(1)}
\right]
+ 8\dot{\tau}\Thz \,\Theta_0^{(1)} \Theta^{(0)}_0\,(Y_1-Y)
}
Where we used $\DiffO \Delta n_0^{(0)}-\delta^{(0)}_{\gamma, 0}\,Y= 4\Theta^{(0)}_0 (Y_1-Y)$. The first group of terms is again solved as before, but the last term remains out of equilibrium. It neither adds energy nor photon number to the spectrum, but just changes the spectral shape. This means that even if the average spectrum has reached a thermal spectrum, the anisotropy spectrum is distorted unless photon production and redistribution are very efficient. We remark that this conclusion does not change when adding back the omitted stimulated scattering term, $2\Thz \DiffO^* \Delta n^{(0)}_0 \Delta n^{(1)}_0$.

\subsection{Photon production in the $\mu$-era}
\label{app:photon_production_terms}

A pure $\mu$-distortion spectrum, $M$, does not balance emission and absorption at low frequencies, leading to net photon production in the $\mu$-era and slow evolution of the $\mu$-parameter at high frequencies.
However, if we insert $\Delta n_0=G(x)\,\Theta_{0}+M(x)\,\mu_{0}$ into the collision term and compute the photon production rate, we obtain a divergent result at low frequencies. We thus need to find a modified solution for the distortion that fixes this problem.

\subsubsection{Zeroth order treatment of photon production}
\label{app:photon_production_terms_zeroth}
If we consider the photon evolution equation at zeroth order, with Eq.~\eqref{app:zeroth_CS} and \eqref{eq:em_zeroth} we have
\bealf{
\label{app:zeroth_CS_DC_photon}
\frac{\partial \Delta n_0^{(0)}}{\partial \tau}
&\approx
\Thz\left[\Theta_{\rm eq}^{(0)}Y+\KompO \Delta n_0^{(0)}\right]
- \frac{\Lambda(x, \Thz)}{x^2}\,\frac{\nbb}{G}\left[\Delta n^{(0)}_0-
\Theta^{(0)}_{\rm eq}\,G\right],
}
where we used $(1-\expf{-x})/x=\expf{-x}(\expf{x}-1)/x=\nbb/G$ and assumed that the heating process led to a non-zero initial chemical potential and temperature shift at the initial time. Clearly, a spectrum $\Delta n^{(0)}_0=\Theta^{(0)}_{\rm eq}\,G$ is a stationary solution of the problem. The strategy is now to assume quasi-stationary evolution of the spectrum, which means the left hand side is set to zero, $\partial_\tau \Delta n_0^{(0)}\approx 0$.
Without emission and absorption terms we know that the spectrum will not change in this case and simply be given by 
$\Delta n_0^{(0)}=G(x)\,\Theta_{0}^{(0)}+M(x)\,\mu_{0}^{(0)}$ with $\Theta_{0}^{(0)}$ and $\mu_{0}^{(0)}$ fixed by the initial condition. However, as already mentioned, at low frequency this does not solve the equation above. Energetically the required correction does not matter much, which means we can still use $\Theta_{\rm eq}^{(0)}\approx \Theta_{0}^{(0)}+\eta_M\,\mu_{0}^{(0)}$.

To obtain the correction to the spectrum, we take the low-frequency limit of Eq.~\eqref{app:zeroth_CS_DC_photon}, essentially thinking of the problem as a high-frequency ($\leftrightarrow$ energy) evolution and low-frequency ($\leftrightarrow$ photon number) evolution \citep[e.g., see][for more details]{Chluba2014}. 
The characteristic scale in the problem is the {\it critical frequency}, $\xc\ll 1$, which is defined by the balance between photon emission processes and Compton scattering \citep{Sunyaev1970mu, Danese1982, Hu1993, Chluba2005, Khatri2012b, Chluba2014}. This frequency is determined (at leading order) by the implicit equation $\Lambda(\xc, \Thz)\approx \Thz \xc^2$, which assumes that the emission coefficient, $\Lambda(\xc, \Thz)$, is a slowly varying function of $x$. In the DC-era, we have $\xc\approx \pot{8.6}{-4}\sqrt{(1+z)/\pot{2}{6}}$, assuming the standard CMB temperature for $\Tz$. The idea is now to write the Ansatz $\Delta n_0^{(0)}=G(x)\,\Theta_{0}^{(0)} g(x)+M(x)\,\mu_{0}^{(0)} m(x)$ with frequency-dependent correction functions $g(x)$ and $m(x)$. At $x\gg \xc$ one has $g(x)\simeq m(x) \simeq 1$. It is therefore useful to consider $g(x)$ and $m(x)$ as functions of $\xi=\xc/x$ instead of $x$. Starting from Eq.~\eqref{app:zeroth_CS_DC_photon}, multiplying it by $-\xc^2/[\Thz\xi^4]$ (this is the leading order dependence) and replacing $x\rightarrow \xc/\xi$, with the Ansatz $\Delta n_0^{(0)}=G(\xc/\xi)\,\Theta_{0}^{(0)} g(\xi)+M(\xc/\xi)\,\mu_{0}^{(0)} m(\xi)$ we then find
\bealf{
\label{app:zeroth_CS_DC_photon_QS_II}
0
&\approx
\mu_{0}^{(0)} \left[\partial_\xi^2  m(\xi)-m(\xi) \right]
+\xc \mu_{0}^{(0)} \left\{\frac{1}{2}\frac{m(\xi)}{\xi}+\eta_M\left[\frac{2-\xi^2}{\xi^3}-\left(\frac{\partial^2}{\partial \xi^2}  \frac{m(\xi)}{\xi}-\frac{m(\xi)}{\xi}\right)\right]\right\}
\nonumber \\
&\qquad \qquad +\xc \Theta_{0}^{(0)}\,\left\{ \frac{2-\xi^2}{\xi^3}[1-g(\xi)]+\frac{4}{\xi^2}\,\partial_\xi g(\xi)\right\}.
}
up to first order in $\xc\ll 1$. Neglecting the terms $\propto \xc$, we have the physical solutions $m(\xi)=\exp(-\xi)$ and $g(\xi)=1$, which together gives
\bealf{
G^{\rm mod}(x)&\approx G(x), \qquad M^{\rm mod}(x)\approx M(x)\,\expf{-\xc/x}
}
for the modified temperature and $\mu$-distortion spectrum. The goal is now use this to obtain an expression for the photon production rate at lowest order in $\xc$. Integrating the collision term in terms of photon number\footnote{There is no contribution from the scattering terms.}, we find
\bealf{
\label{app:zeroth_N_dto}
\frac{1}{N_z}\,\frac{\partial N_{\rm c}^{(0)}}{\partial \tau}
&= -\frac{\int x^2 \id x\, \frac{\Lambda(x, \Thz)}{x^2}\,\frac{\nbb}{G}\left[\Delta n^{(0)}_0-
\Theta^{(0)}_{\rm eq}\,G\right]}{\int x^2 \nbb(x)\id x}
\approx-\frac{\Thz}{N_{\nbb}}
\int_0^\infty \xc^2\left[\Delta n_0^{(0)} -  \,\Theta^{(0)}_{\rm eq}\,G\right]\!\id x
\nonumber\\[1mm]
&=-\frac{\Thz}{N_{\nbb}}
\int_0^\infty \xc^2\left[
G\,\Theta^{(0)}_0  +M\,m\,\mu_0^{(0)}
- G(\Theta^{(0)}_0+\eta_M\,\mu_0^{(0)})\right]\id x
\nonumber\\[1mm]
&\approx \frac{\Thz}{N_{\nbb}}
\int_0^\infty \frac{\xc^2}{x^2}\,\mu_0^{(0)}\,m(x)\id x
\nonumber\\[1mm]
&= \frac{\Thz\,\mu_{0}^{(0)}}{N_{\nbb}}\,
\int_0^\infty \frac{\xc^2}{x^2}\,\expf{-\xc/x}\,\id x
=
\left(\frac{3}{4\,\gamma_\rho}\right) \,\gamma_N\,\Thz\,\xc\,\mu_{0}^{(0)}.
}
from the collision term. This remarkable result is the key to understanding the rate of conversion from $\mu_{0}^{(0)}\rightarrow \Theta^{(0)}_0$.
Indeed, the reduction of the chemical potential amplitude is given by $\partial_t \mu_0^{(0)}\big|_{\rm em/abs}\approx - \gamma_N\,\dot{\tau}\Thz\,\xc\,\mu_{0}^{(0)}$ \citep{Chluba2014}, which we use in main text.
We highlight again that without the low-frequency modification to the spectrum by the factor $\expf{-\xc/x}$ one would not have obtained a finite photon production rate. At low frequencies, the distortion vanishes and the spectrum is in full equilibrium with the electrons, stopping any emission and absorption.

We note that, while for the temperature term $\propto \Theta^{(0)}_0$ an exact cancellation occurred, in the final steps we had to drop the term $\eta_M \mu_0^{(0)}$, since it still diverges logarithmically towards $x\rightarrow 0$. This is because we did not include higher orders in $\xc$.
We can in principle use Eq.~\eqref{app:zeroth_CS_DC_photon_QS_II} to obtain correction to the distortion spectra at first order in $\xc$. However, the procedure becomes quite involved requiring additional normalisation conditions and also the including of time-dependent corrections \citep[see][]{Chluba2014}. This is beyond the scope of this work and left to a future publication. 

We also note that one can in principle evaluate the exact integral $\int \xc^2 M(x) m(x) \id x$ numerically. This yields the correction
$\xc \rightarrow \xc (1-1.65 \xc^{0.88})$, which indicates that photon production is a little slower than in the soft photon limit. However, in terms of perturbations in $\xc$ this is not a consistent treatment and other corrections are expected to have a similar level. We therefore do not recommend adding these corrections until a more complete treatment of the problem is attempted.

\subsubsection{First order treatment of photon production}
\label{app:photon_production_terms_first}
We can now repeat the same computation but at first order in perturbations. The emission terms at first order in perturbations is given in Eq.~\eqref{eq:em_first}. Remembering that $\Delta n^{(0)}_0\approx - \mu_0^{(0)}\,m(x)/x^2$ gave the leading order term at zeroth order and checking the coefficients of all functions, keeping only leading order terms, we then find 
\begin{align}
\nonumber
\frac{\partial n^{(1)}_0}{\partial \tau}\Bigg|^{\rm low}_{\rm em/abs}
&\approx 
\frac{\Thz \xc^2}{x^2} \left[\mu_0^{(1)}\,\frac{m^{(1)}(x)}{x^2}
+\left(\delta^{(1)}_{\rm b}+\Psi^{(1)}+\Theta^{(1)}_0  
\left[
\frac{\partial \ln \Lambda}{\partial\ln\Thg}\Bigg|_{\Thz}+\frac{\partial \ln \Lambda}{\partial\ln\The}\Bigg|_{\Thz}-1
\right]\right)\,\mu_0^{(0)}\,\frac{m^{(0)}(x)}{x^2}\right].
\end{align}
Assuming $\Lambda\propto \Thg^2$, we have $\frac{\partial \ln \Lambda}{\partial\ln\Thg}=2$. Realising that $Y_1-Y\approx 3/[2x]$ and $Y\approx -2/x$ at $x\ll 1$, we can also collect all leading order Compton terms in Eq.~\eqref{app:first_order_CS} as 
\bealf{
\frac{\partial \Delta n_0^{(1)}}{\partial \tau}\Bigg|^{\rm low}_{\rm K}
&\approx
\Thz \left\{
\KompO^{\rm low} \Delta n_0^{(1)}
+
\left[\delta_{\rm b}^{(1)}+\Psi^{(1)}\right] \KompO^{\rm low} \Delta n_0^{(0)}
+\Theta_0^{(1)} \DiffO \Delta n_0^{(0)}
+ 2\DiffO^* \Delta n^{(0)}_0 \Delta n^{(1)}_0\right\}
\nonumber
}
with $\KompO^{\rm low}=x^{-2} \partial_x x^4 [\partial_x+2/x]\equiv x^{-2} \partial_x x^2 \partial_x x^2$.
We added the term $2\DiffO^* \Delta n^{(0)}_0 \Delta n^{(1)}_0\approx 2\Theta^{(1)}_0\DiffO^* \Delta n^{(0)}_0 G(x)$, which naturally cancels some of the logarithmic corrections at low frequencies. However, in the overall evolution at high frequencies it should not be as important.

From the first order computation it is clear that the term $\propto \delta_{\rm b}^{(1)}+\Psi^{(1)}$ will be canceled by the corresponding zeroth order emission term.
Inserting $\Delta n^{(i)}_0\approx - \mu_0^{(i)}\,m^{(i)}(x)/x^2$ and adding the emission terms we then obtain
\begin{align}
\nonumber
0&\approx 
\mu_0^{(1)}\,\KompO^{\rm low} \frac{m^{(1)}(x)}{x^2}
+\Theta^{(1)}_0\,\mu_0^{(0)}\,\left[\DiffO \frac{m^{(0)}(x)}{x^2}
+\DiffstarO \frac{m^{(0)}(x)}{x^3}\right]
-
\frac{\xc^2}{x^4} \left[\mu_0^{(1)}\,m^{(1)}(x)
+\Theta^{(1)}_0\,\mu_0^{(0)}\,m^{(0)}(x)\right].
\end{align}
after dividing through by $-\Thz$. Transforming to $\xi=\xc/x$ with $\partial_x = -(\xc/x^2) \partial_\xi$ we then have $\KompO^{\rm low}=(\xc^2/x^4) \,\partial_\xi^2 x^2$, $\DiffO x^{-2}=(\xc^2/x^4) \,\partial_\xi \xi^{-2}\partial_\xi  \xi^2$ and $\DiffstarO x^{-3}=-(\xc^2/x^4) \,\partial_\xi \xi^{-1}$ such that $\DiffO m^{(0)}(x)/x^{2}=(\xc^2/x^4) [1-2/\xi-2/\xi^2] m^{(0)}(\xi)$ and 
$\DiffstarO m^{(0)}(x)/x^{3}=(\xc^2/x^4) [2/\xi+2/\xi^2] m^{(0)}(\xi)$
\begin{align}
\nonumber
0&\approx 
\mu_0^{(1)}\left[\partial_\xi^2 m^{(1)}(\xi)
-m^{(1)}(\xi)\right]
+\Theta^{(1)}_0\,\mu_0^{(0)}\left[\partial_\xi \xi^{-2}\partial_\xi \xi^2 m^{(0)}(\xi)-
\partial_\xi \xi^{-1}\,m^{(0)}(\xi)
-m^{(0)}(\xi)
\right]
\nonumber\\
&\approx 
\mu_0^{(1)}\,m^{(0)}(\xi)\left[\partial_\xi^2 f
-2\partial_\xi f \right].
\end{align}
where in the last line we made the Ansatz $m^{(1)}(\xi)=m^{(0)}(\xi)\,f(\xi)$. We then obtain $f=1$ as the main physical solution. This means that the first order perturbed distortion spectrum is identical to that of the zeroth order and highlights that the Comptonisation and emission correction terms cancel each other.
We can therefore directly write the full photon production term as
\bealf{
\label{app:first_N_dto}
\frac{1}{N_z}\,\frac{\partial N_{\rm c}^{(1)}}{\partial \tau}
&\approx 
\left(\frac{3}{4\,\gamma_\rho}\right) \,\gamma_N\,\Thz\,\xc\,\left[\mu_{0}^{(1)} + \left(\delta^{(1)}_{\rm b}+\Psi^{(1)}+\Theta^{(1)}_0
\right) \mu_{0}^{(0)}\right].
}
In addition to the expected term $-\gamma_N\,\dot{\tau}\Thz\,\xc\,\mu_{0}^{(1)}$, we therefore have the correction $-\gamma_N\,\dot{\tau}\Thz\,\xc \Big(\delta^{(1)}_{\rm b}+\Psi^{(1)}+\Theta^{(1)}_0
\Big) \, \mu_{0}^{(0)}$, yielding a total reduction rate of
\bealf{
\label{app:first_mu_dot}
\partial_t \mu^{(1)}_0\approx -\gamma_N\,\dot{\tau}\Thz\,\xc\,\left[\mu_{0}^{(1)} + \left(\delta^{(1)}_{\rm b}+\Psi^{(1)}+\Theta^{(1)}_0
\right) \mu_{0}^{(0)}\right].
}
Following our approach for the zeroth order evolution, this can be used to model the redistribution of energy between $\mu^{(1)}_0$ and $\Theta^{(1)}_0$.

In the $\mu$-era, we have a photon production rate $\simeq \Thg \xc\propto \Thg^{3/2}$, where the local photon temperature matters. If the temperature changes, then both photon diffusion and DC/BR emission find a new equilibrium, naively yielding a perturbation to the photon production term $\simeq [\Thg \xc(\Thg)]^{(1)} \propto \Thz \xc(\Thz)\,[(1+\Theta_0)^{3/2}]^{(1)}\approx \Thz \xc(\Thz) \,(3/2)\,\Theta_0^{(1)}$, as also used in Sect.~\ref{sec:Comp_effects_simplistic}. Why is the coefficient in front of $\Theta^{(1)}_0$ equal to 1? 
The reason is that in comparison to the zeroth order treatment we have three unexpected terms appearing in our quasi-stationary equation. For the emission process, this is the term $-\Thz (\xc^2/x^4) \mu_0^{(0)} m^{(0)}(x)$, while for the Compton contribution they are $\Thz \Theta_0^{(1)} \DiffO \Delta n_0^{(0)}
+ 2\Thz \DiffO^* \Delta n^{(0)}_0 \Delta n^{(1)}_0$.
Neglecting all these terms we have the quasi-stationary equation
\begin{align}
\nonumber
0&\approx 
\mu_0^{(1)}\,\KompO^{\rm low} \frac{m^{(1)}(x)}{x^2}
-
\frac{\xc^2}{x^4} \left[\mu_0^{(1)}\,m^{(1)}(x)
+2\Theta^{(1)}_0\,\mu_0^{(0)}\,m^{(0)}(x)\right].
\end{align}
Carrying out the transformations as above, this then yields the equation
\begin{align}
\nonumber
0&\approx 
\left\{\mu_0^{(1)}\left[\partial_\xi^2 f
-2\partial_\xi f \right]
-2\Theta^{(1)}_0\,\mu_0^{(0)}\right\}
m^{(0)}(\xi),
\end{align}
which has the physical solution $f=1-\Theta^{(1)}_0\,\mu_0^{(0)}\,\xi/\mu_0^{(1)}$ or $\mu_0^{(1)} m_0^{(1)}(\xi)\approx (\mu_0^{(1)}-\Theta^{(1)}_0\,\mu_0^{(0)}\,\xi)\,m_0^{(0)}(\xi)$. Evaluating the photon production term with this correction, we find a contribution $-\Theta^{(1)}_0\mu_0^{(0)}$ from the terms $\mu_0^{(1)}\,m^{(1)}(\xi)$, which put together again yields the correct result, 
$-\Theta^{(1)}_0\,\mu_0^{(0)}+2\Theta^{(1)}_0\,\mu_0^{(0)}=\Theta^{(1)}_0\,\mu_0^{(0)}$.

If we now only neglect the Compton terms $\Thz \Theta_0^{(1)} \DiffO \Delta n_0^{(0)}
+ 2\Thz \DiffO^* \Delta n^{(0)}_0 \Delta n^{(1)}_0$, we find
\begin{align}
\nonumber
0&\approx 
\left\{\mu_0^{(1)}\left[\partial_\xi^2 f
-2\partial_\xi f \right]
-\Theta^{(1)}_0\,\mu_0^{(0)}\right\}
m^{(0)}(\xi).
\end{align}
yielding $\mu_0^{(1)} m_0^{(1)}(\xi)\approx (\mu_0^{(1)}-\frac{1}{2}\Theta^{(1)}_0\,\mu_0^{(0)}\,\xi)\,m_0^{(0)}(\xi)$ and hence a contribution $-\Theta^{(1)}_0\,\mu_0^{(0)}/2$ from the production integral over $\mu_0^{(1)}\,m^{(1)}(\xi)$. Put together this then gives 
$-\Theta^{(1)}_0\,\mu_0^{(0)}/2+\Theta^{(1)}_0\,\mu_0^{(0)}=(1/2)\,\Theta^{(1)}_0\,\mu_0^{(0)}$, which now is short of the expected result. Indeed, if we would only have used $[\xc]^{(1)}\simeq[(1+\Theta_0)^{1/2}]^{(1)}$, this would have been the result, highlighting the link to the emission process only.\footnote{The extra factor $\Thg$ in the photon production term $\propto \Thg \xc$ comes from Compton scattering only.}

If we only neglect the extra photon emission term, $-\Thz (\xc^2/x^4) \mu_0^{(0)} m^{(0)}(x)$, we {\it again} obtain 
\begin{align}
\nonumber
0&\approx 
\left\{\mu_0^{(1)}\left[\partial_\xi^2 f
-2\partial_\xi f \right]
-\Theta^{(1)}_0\,\mu_0^{(0)}\right\}
m^{(0)}(\xi).
\end{align}
and hence a contribution $-\Theta^{(1)}_0\,\mu_0^{(0)}/2$ from the production integral over $\mu_0^{(1)}\,m^{(1)}(\xi)$. But now adding things together we have $-\Theta^{(1)}_0\,\mu_0^{(0)}/2+2\Theta^{(1)}_0\,\mu_0^{(0)}=(3/2)\,\Theta^{(1)}_0\,\mu_0^{(0)}$. This suggests that the naive estimate does not capture the full effect of changing the DC emissivity with temperature. 

While this is somewhat satisfying, we note that the required effect stems from a higher order correction caused by the absorption term $\propto - \Delta n_0 \,(\expf{x\,\Thz/\The}-1)$ in Eq.~\eqref{eq:gen_emissionterm}, which does not contribute at zeroth order, and hence can also not be captured by perturbing the zeroth order equation. 
%
%
In reality, additional corrections can be expected since the frequency dependence of the DC emissivity, $\Lambda$, may also modify matters. This was highlighted before as part of the analytic computations for the distortion visibility function \citep{Chluba2005, Chluba2014}. Overall this shows that the precise scaling may depend on which approximation for the photon emissivity is indeed used. However, a more detailed discussion is beyond the scope of this work. 

\newpage 

\section{Details of the derivation for the line-of-sight integral solution.}
\label{app:LOS}
To simplify Eq.~\eqref{eq:evol_Yi_1st_ord_Fourier_I} and obtain an expression for $\tilde{\vek{y}}^{(1)}_\ell(\eta_f, k)$ at the conformal time $\eta_f$, we follow the standard steps \citep{CMBFAST}.
Defining the Thomson optical depth as $\tau=\int_0^\eta \tau'(\eta') \id \eta'$, it then follows
\begin{align}
\partial_\eta \vek{y}^{(1)}
+\i k \chi \,\vek{y}^{(1)}
+\tau' \vek{y}^{(1)}=\expf{-(\i k\chi \eta+\tau)}\partial_\eta\left[\expf{\i k\chi \eta+\tau}\,\vek{y}^{(1)}\right].
\end{align}
We can therefore formally integrate Eq.~\eqref{eq:evol_Yi_1st_ord_Fourier_I} \citep{CMBFAST}
\begin{align}
\label{eq:formal_sol}
\vek{y}^{(1)}(\eta_f, \chi, k)=
\int_0^{\eta_f} \id \eta \, 
\expf{-\i k \chi \Delta \eta -\tau_{\rm b}}\,
\vek{S}_\text{LOS}(\eta, \chi, k)
\end{align}
to obtain the solution at the final time $\eta_f$. 
Here we defined $\tau_{\rm b}=\tau(\eta_f)-\tau(\eta)$ and $\Delta \eta=\eta_f-\eta$ for convenience. Note that now $\partial_\eta \tau_{\rm b}=-\tau'$.

In the source term, three types angular dependencies appear:
\bsub
\label{eq:evol_Yi_1st_ord_Fourier_II}
\begin{align}
\vek{S}_\text{LOS, 0} &=
    \tau' \tilde{\vek{y}}_0^{(1)}
    -\vek{b}^{(0)}_0 \frac{\partial \tilde{\Phi}^{(1)}}{\partial \eta}
+\frac{{\vek{Q}'}^{(1)}}{4},
\\
\vek{S}^{{\rm therm}}_\text{LOS, 0} &=
\tau'\Thz\left[M_{\rm K}\,\tilde{\vek{y}}^{(1)}_0\!+\!\vek{D}^{(1)}_0
\!+\!\left[\tilde{\delta}_{\rm b}^{(1)}+\tilde{\Psi}_{\rm b}^{(1)}\right]\left(M_{\rm K}\,\vek{y}^{(0)}_0\!+\!\vek{D}^{(0)}_0\right)
\!+\!\tilde{\Theta}^{(1)}_0\,\vek{D}^{(0)}_0\right]
\\
\vek{S}_\text{LOS, 1} &\equiv 
    -\vek{b}^{(0)}_0 \i k \chi\Psi^{(1)} 
+\tau'\beta^{(1)}\chi\,\vek{b}^{(0)}_0
=-\i k \chi\left[\tilde{\Psi}^{(1)} 
+\frac{\tau'}{k}\,\tilde{\beta}^{(1)} \right]\vek{b}^{(0)}_0,
\\
\vek{S}_\text{LOS, 2} &\equiv 
\frac{\tau' }{10}\,\vek{y}_2^{(1)}= -\frac{\tau' }{2}\,\tilde{\vek{y}}_2^{(1)}\,P_2(\chi).
\end{align}
\esub
Here we separated thermalisation terms $\propto \tau'\Thz$ from other sources.
Inserting this back into Eq.~\eqref{eq:formal_sol} and carrying out the Legendre transform, we then have
\bsub
\label{eq:formal_sol_Leg}
\begin{align}
    \tilde{\vek{y}}^{(1)}_\ell(\eta_f, k) &=
    \frac{\i^\ell}{2}\int P_\ell(\chi)\,\vek{y}^{(1)}(\eta_f, \chi, k)\id \chi
    = \int_0^{\eta_f} \id \eta \,g(\eta)\, \tilde{\mathcal{\vek{S}}}_\ell(\eta, \eta_f, k),
    \\
\label{eq:formal_sol_Leg_b}
\tilde{\mathcal{\vek{S}}}_\ell(\eta, \eta_f, k)&=\frac{\i^\ell}{2}\int P_\ell(\chi)\,\expf{-\i k \chi \Delta \eta}\,\frac{\vek{S}_\text{LOS}(\eta, \chi, k)}{\tau'}\id \chi,  
\end{align}
\esub
where we introduced the visibility function, $g(\eta)=\tau'\,\expf{-\tau_{\rm b}}=\partial_\eta \expf{-\tau_{\rm b}}$. Using the identity \citep{Hu1997}
\begin{align}
\expf{-\i k \chi \Delta \eta}&=\sum_\ell (-\i)^\ell (2\ell+1) j_\ell[k\Delta \eta]\,P_\ell(\chi),  
\end{align}
in terms of spherical bessel function, $j_\ell(x)$, we then encounter the following cases
\bsub
\label{eq:formal_sol_Leg_app}
\begin{align}
&\frac{\i^\ell}{2}\int P_\ell(\chi)\,\expf{-\i k \chi \Delta \eta}\id \chi=j_\ell(k\Delta \eta),
\\
&\frac{\i^\ell}{2}\int P_\ell(\chi)\,(-\i k\chi)\,\expf{-\i k \chi \Delta \eta}\id \chi
=
\partial_{\Delta\eta} j_\ell(k\Delta\eta)
=
k\,
\left[\frac{\ell}{2\ell+1} j_{\ell-1}(k\Delta\eta)-\frac{\ell+1}{2\ell+1}j_{\ell+1}(k\Delta\eta)\right]
, 
\\
&\frac{\i^\ell}{2}\int P_\ell(\chi)\,[-P_2(\chi)]\,\expf{-\i k \Delta\chi \eta}\id \chi=
\frac{j_\ell(k\Delta\eta)}{2}+\frac{3}{2}\,
\partial^2_{k\Delta \eta} j_\ell(k\Delta\eta),  
\end{align}
\esub
in Eq.~\eqref{eq:formal_sol_Leg_b}. Putting things together we then find $\tilde{\mathcal{\vek{S}}}_\ell(\eta, \eta_f, k)$ as given in Eq.~\eqref{eq:formal_sol_Leg_fin}.\footnote{As an intermediate step we used that  $\int g(\eta)\,\frac{k}{\tau'}\,\tilde{\Psi}^{(1)}\,j'_\ell(k\eta)\,\id \eta=\int g(\eta)\,\frac{1}{\tau'}\,[\partial_\eta\tilde{\Psi}^{(1)}] j_\ell(k\eta)\id \eta+\int g(\eta) \tilde{\Psi}^{(1)}\,j_\ell(k\eta)\id\eta $ since $k\tilde{\Psi}^{(1)}\,j'_\ell(k\Delta\eta)=[\partial_\eta\tilde{\Psi}^{(1)}] j_\ell(k\Delta\eta)-\partial_\eta [\tilde{\Psi}^{(1)}\,j_\ell(k\Delta \eta)]$.}


\end{document}